\documentclass[lettersize,journal]{IEEEtran}
\usepackage{amsmath,amssymb,amsfonts}
\usepackage{algorithmic}
\usepackage{algorithm}
\usepackage{array}

\usepackage[caption=false,font=normalsize,labelfont=sf,textfont=sf]{subfig}
\usepackage{textcomp}
\usepackage{stfloats}
\usepackage{url}
\usepackage{verbatim}
\usepackage{graphicx}
\usepackage{cite}
\usepackage{tabularray}
\usepackage{tabularx}
\UseTblrLibrary{siunitx}
\usepackage{booktabs}
\usepackage{colortbl}
\usepackage{multirow}
\usepackage[normalem]{ulem}
\usepackage{diagbox}
\usepackage[version=4]{mhchem}
\usepackage[dvipsnames]{xcolor}
\usepackage[colorlinks=true, citecolor=ForestGreen, linkcolor=Orchid, urlcolor=NavyBlue]{hyperref}

\usepackage[switch]{lineno}

\begin{document}

\title{Zero-Shot Image Denoising for High-Resolution Electron Microscopy}

\author{Xuanyu Tian,~\IEEEmembership{Graduate Student Member,~IEEE,}
Zhuoya Dong, 
Xiyue Lin,
Yue Gao,~\IEEEmembership{Senior~Member,~IEEE,}
Hongjiang Wei,~\IEEEmembership{Member,~IEEE,}
Yanhang Ma,
Jingyi Yu,~\IEEEmembership{Fellow,~IEEE,}
and Yuyao Zhang,~\IEEEmembership{Member,~IEEE}
\thanks{ (\textit{Corresponding authors: Jingyi Yu; Yuyao Zhang})}
\thanks{
Xuanyu Tian is with School of Information Science and Technology, ShanghaiTech University, 201210, Shanghai, China and Lingang Laboratory, 20031, Shanghai, China (e-mail: tianxy@shanghaitech.edu.cn)
}
\thanks{Xiyue Lin, Jingyi Yu and Yuyao Zhang are with School of Information Science and Technology, ShanghaiTech University, 201210, Shanghai, China (e-mail: linxy2022@shanghaitech.edu.cn; yujingyi@shanghaitech.edu.cn; zhangyy8@shanghaitech.edu.cn)
}
\thanks{
Zhuoya Dong and Yanhang Ma are with School of Physical Science and Technology, ShanghaiTech University, 201210, Shanghai, China (e-mail: dongzhy@shanghaitech.edu.cn; mayh2@shanghaitech.edu.cn)
}
\thanks{
Yue Gao is with the BNRist, THUIBCS, KLISS, BLBCI, School of Software, Tsinghua University, Beijing 100084, China (e-mail: gaoyue@tsinghua.edu.cn).
}
\thanks{
Hongjiang Wei is with the School of Biomedical Engineering and Institute of Medical Robotics, Shanghai Jiao Tong University, 200127 Shanghai, China (e-mail: hongjiang.wei@sjtu.edu.cn).
}
\thanks{The code for this work is availble at \url{https://github.com/MeijiTian/ZS-Denoisier-HREM} }
}

\markboth{Journal of \LaTeX\ Class Files,~Vol.~14, No.~8, August~2021}%
{Shell \MakeLowercase{\textit{et al.}}: A Sample Article Using IEEEtran.cls for IEEE Journals}


\maketitle
\begin{abstract}
High-resolution electron microscopy (HREM) imaging technique is a powerful tool for directly visualizing a broad range of materials in real-space.
However, it faces challenges in denoising due to ultra-low signal-to-noise ratio (SNR) and scarce data availability.
In this work, we propose Noise2SR, a zero-shot self-supervised learning (ZS-SSL) denoising framework for HREM.
Within our framework, we propose a super-resolution (SR) based self-supervised training strategy, incorporating the Random Sub-sampler module. 
The Random Sub-sampler is designed to generate approximate infinite noisy pairs from a single noisy image, serving as an effective data augmentation in zero-shot denoising.
Noise2SR trains the network with paired noisy images of different resolutions, which is conducted via SR strategy.
The SR-based training facilitates the network adopting more pixels for supervision, and the random sub-sampling helps compel the network to learn continuous signals enhancing the robustness.
Meanwhile, we mitigate the uncertainty caused by random-sampling by adopting minimum mean squared error (MMSE) estimation for the denoised results.
With the distinctive integration of training strategy and proposed designs, Noise2SR can achieve superior denoising performance using a single noisy HREM image.
We evaluate the performance of Noise2SR in both simulated and real HREM denoising tasks.
It outperforms state-of-the-art ZS-SSL methods and achieves comparable denoising performance with supervised methods. 
The success of Noise2SR suggests its potential for improving the SNR of images in material imaging domains.

\end{abstract}


\section{Introduction}
\IEEEPARstart{{H}}{{igh}}-resolution electron microscopy (HREM) imaging~\cite{o1978computed,spence1981experimental,kisielowski2008detection} is an indispensable tool in the fields of materials science and nanotechnology.
HREM enables direct visualization of structures at the atomic level through interactions between the sample and high-energy electrons. 
{However}, HREM is inevitably susceptible to noise {due to inherent properties of electron beams and detection process \textit{etc}.}
{For example, the low-dose conditions are often applied to imaging electron-beam sensitive materials to minimize the damage from the electrons~\cite{kato2004reducing,jiang2015electron}.}
{Capuring the dynamic events at kilohertz frame rates with direct electronic detection systems~\cite{faruqi2018direct, ercius20204d}, the image is affected by severe shot noise due to the shortened exposure time.}
{Improving signal-to-noise (SNR)} is critical for HREM to enhance image quality and facilitate accurate information extraction. 

Recently, data-driven methods based on deep learning (DL)~\cite{zhang2017beyond,tai2017memnet,zhang2018ffdnet} have obtained favorable performance compared to conventional methods in image denoising.
However, applying supervised DL denoising methods to electron microscopy (EM) images is challenging due to the lack of paired noisy-clean image datasets.
In the absence of ground-truth images, several self-supervised image denoising methods\cite{batson2019noise2self, krull2019noise2void, laine2019high, cha2019fully,  quan2020self2self, xie2020noise2same, byun2021fbi, moran2020noisier2noise, huang2022neighbor2neighbor,pan2023random}
have been proposed. 
Some works~\cite{krull2019noise2void, batson2019noise2self, laine2019high, byun2021fbi} utilize blind-spot networks (BSNs) to prevent identical mappings in self-supervised learning. 
BSNs predict the clean value of a noisy pixel without using the information from that pixel itself. To build BSN, Noise2Void~\cite{krull2019noise2void} and Noise2Self~\cite{batson2019noise2self} initially employ a masked-based strategy, while subsequent works~\cite{cha2019fully, laine2019high, wu2020unpaired, byun2021fbi} propose tailored network designs.
Huang \textit{et al.} proposes  Neighbor2Neighbor (NB2NB)~\cite{huang2022neighbor2neighbor} to generate paired noisy images from subsampling paired neighbor pixels for self-supervised training.
While NB2NB and its variants~\cite{lequyer2022fast,mansour2023zero} fail to address the intensity gap issue between neighbor pixels, resulting in relatively poor performance in low-SNR scenarios.
Additionally, some works~\cite{moran2020noisier2noise, pang2021recorrupted} introduce explicit noise modeling to generate training paired noisy images by adding synthetic noise.
However, in HREM imaging, the noise distribution of real HREM is unknown and hard to estimate.
Furthermore, the scarcity of HREM data poses a significant challenge for these methods since their performance tends to degrade with insufficient data training~\cite{quan2020self2self, mansour2023zero}. 

In this paper, we propose an efficient zero-shot self-supervised denoising framework for HREM images named Noise2SR.
We propose to train a denoising network with paired noisy images with \textit{different resolutions}, which is conducted via super-resolution (SR) strategy.
Inspired by NB2NB~\cite{huang2022neighbor2neighbor}, we introduce the Random Sub-sampler module to generate sub-sampled noisy images that form a noisy pair with the original noisy image.
Unlike NB2NB, we utilize paired noisy images with different resolutions for training.
We indicated that the sub-sampled noisy image and the original noisy image have a consistent underlying clean image.
Meanwhile, we provide theoretical proof that the Noise2SR training scheme is statically equivalent to using a clean image for supervision.
Combined with the SR-based training strategy, we address the coordinate mismatch and intensity gap issues present in paired neighbor pixels in NB2NB. 
The proposed Random Sub-sampler helps to break the noise correlation of real-world and serves as an effective data augmentation for enhancing the zero-shot image denoising performance.
We adopt minimum mean squared error (MMSE) estimation to mitigate the uncertainty caused by random sub-sampling and further enhance the denoising performance.
With the distinctive integration of SR training strategy, Random sub-sampling and MMSE estimation, Noise2SR exhibits significant performance for single image denoising, particularly in scenarios with extremely low signal-to-noise ratios (SNR) such as HREM. 
We conduct a series of experiments on both simulated and real HREM images to demonstrate the effectiveness and superiority of the proposed Noise2SR framework for HREM image denoising.
The main contributions of this work can be summarized as follows:
\begin{enumerate}
    \item
    We proposed Noise2SR, which efficiently improves the signal-to-noise ratio (SNR) of single HREM images without involving any external dataset. To the best of our knowledge, Noise2SR could be one of the first zero-shot self-supervised denoising methods for HREM images.
    \item 
    We propose a novel self-supervised training scheme incorporating SR strategies without noise model assumptions and can be combined with any network or framework.
    \item We propose the Random Sub-sampler serves as an effective data augmentation in network training and incorporates MMSE estimation to effectively produce reliable denoised results in ultra-low SNR scenarios.
    \item Incorporating the training scheme and designs,  our method performs very favorably against state-of-the-art self-supervised denoising methods in HREM image denoising. 
\end{enumerate}

The quantitative and qualitative results demonstrate the effectiveness of our methods in enhancing the SNR of simulated HREM images and low-dose HREM images of two electron beam sensitive zeolites.

This work is built upon our previous work~\cite{tian2022noise2sr} and introduces several notable improvements.
Firstly, we extend the previous work to address the denoising of single low-dose HREM data. 
Secondly, we propose an approximate MMSE estimation in the inference stage, which enhances the denoising performance and provides stability in the context of single HREM image denoising. 
Furthermore, we conduct a comprehensive discussion on the proposed Random Sub-sampler module and evaluate its effectiveness within the overall framework.

\section{Related Work}

\subsection{Image Denoising Without Clean Signal Prior}
In recent years, supervised image denoising methods based on deep neural networks \textit{e.g.}, DnCNN~\cite{zhang2017beyond} have achieved great success and outperform conventional image denoiser.
However, the acquisition of aligned noisy-clean images is infeasible and impractical in many scientific imaging applications, such as electron microscopy, which limits the use of supervised deep learning approaches.

Noise2Noise (N2N)~\cite{lehtinen2018noise2noise} is the first work that proposes training a deep denoiser using paired noisy images of the same scene and demonstrates that it is statistically equivalent to supervised learning.
Subsequently, self-supervised denoising methods have been proposed, 
{enabling the denoiser to be trained from individual noisy images without paired noisy images.}
Noise2Void (N2V)~\cite{krull2019noise2void} and Noise2Self (N2S)~\cite{batson2019noise2self} design the masked-based blind-spot network (BSN) for self-supervised learning.
Specifically, the masked-based BSN replaces certain pixels of input noisy images and predicts their value based on neighboring pixels.
However, the replacement strategy cannot entirely prevent the network from learning identical mappings in self-supervised denoising, as pointed out by \cite{xie2020noise2same}. 
Additionally, the masked-based BSNs suffer from limited pixels for supervision, resulting in inefficient training~\cite{laine2019high} and artifacts in the denoised results.
To address these issues, the following works~\cite{laine2019high,wu2020unpaired,cha2019fully,byun2021fbi,lee2022ap} have proposed to design novel BSN architectures {by incorporating centrally masked convolution and dilated convolutions.}
{However, these methods are limited by the large amount of calculation and the inflexible network structure.}
Different from BSN methods, another category of self-supervised learning approaches proposed generating approximate paired noisy images from individual noisy images.
For example, Noisier2Noise~\cite{moran2020noisier2noise} generates paired noisy images by adding additional noise to noisy images, and similar ideas are explored in~\cite{xu2020noisy,pang2021recorrupted}. 
However, these methods require a known noise model of original noisy images to guarantee denoising performance.
Additionally, Neighbor2Neighbor (NB2NB)~\cite{huang2022neighbor2neighbor} proposed to sub-sample the neighbor pixels of the noisy image to generate paired noisy images.
Since a gap exists between the underlying ground truth of paired neighbor sub-sampled images, NB2NB has limited performance in low-SNR scenarios.

\subsection{Advances in Zero-shot Image Denoising}
Zero-shot learning in image denoising refers to training a denoiser using only a single noisy image without any external data.
This approach is particularly valuable in scenarios where training data is scarce, such as in HREM.  
The main challenge is avoiding overfitting, which can significantly reduce the effectiveness of the aforementioned self-supervised denoising methods as training data decreases.
Deep image prior (DIP) pioneered zero-shot image denoising by utilizing the early-stopping strategies in convolutional neural networks (CNNs) training. 
With the prosperity of self-supervised denoising, Self2Self (S2S) trained a zero-shot denoiser that incorporates a dropout regularization with a blind-spot networks framework to mitigate overfitting.
FBI-Denoiser~\cite{byun2021fbi} showed superior performance of zero-shot denoising in Poission-Gaussian noise denoising through careful BSN design. 
Recent studies~\cite{lequyer2022fast,mansour2023zero} explored fast and efficient zero-shot denoisers by introducing a novel sub-sampling strategy or a lightweight network.

\begin{figure*}[!t]
    \centering
    \includegraphics[width=\textwidth]{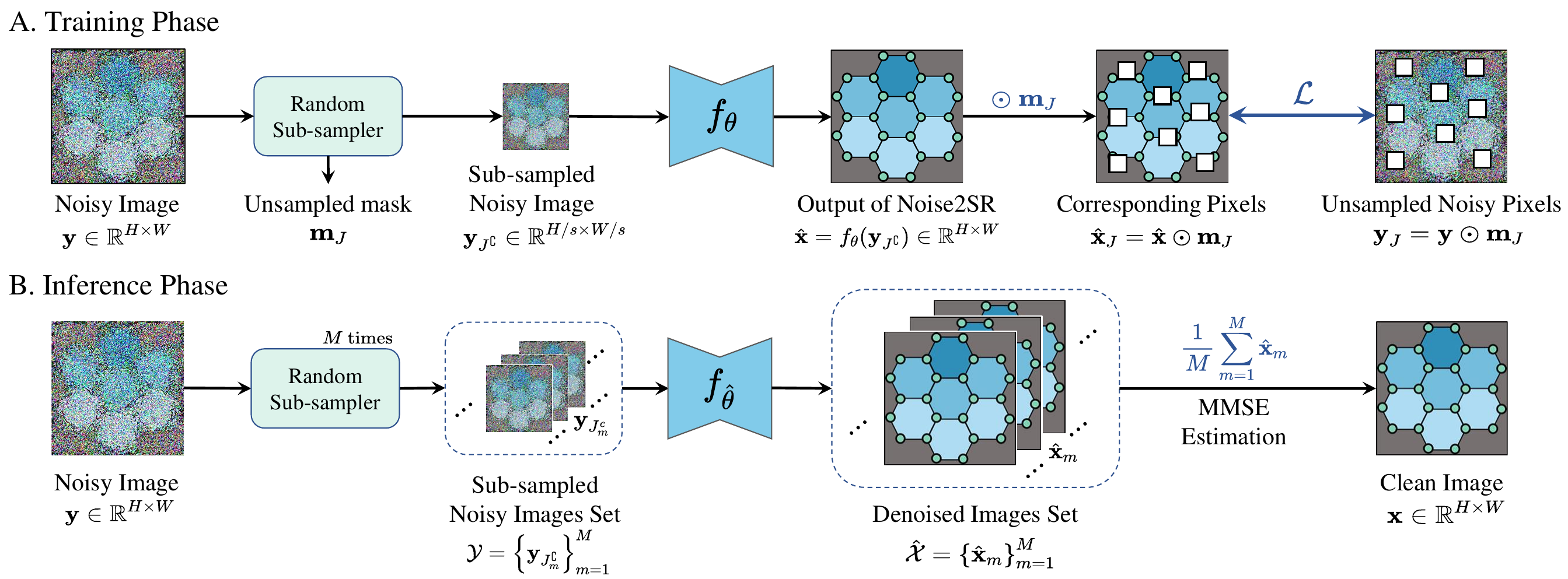}
    \caption{Pipeline of proposed Noise2SR framework. \textbf{A. Training Phase:} First, Random Sub-sampler takes a noisy image $\mathbf{y}$ as input and generates a sub-sampled noisy image $\mathbf{y}_{J^\complement}$ along with corresponding unsampled mask $\mathbf{m}_J$. Then, the network $f_\theta$ takes the sub-sampled image $\mathbf{y}_{J^\complement}$ as input and generates a denoised image of full resolution $f_\theta(\mathbf{y}_{J^\complement})$. The network is optimized by computing the loss on the difference between unsampled noisy pixels $\mathbf{y}_J$ and the output of the network. \textbf{B. Inference Phase:} A sub-sampled noisy set $\mathcal{Y}$ can be obtained by repeatedly sub-sampling a noisy image $\mathbf{y}$ $M$ times using the Random Sub-sampler. Given a sub-sampled noisy set $\small{\mathcal{Y}}$, well-trained network $f_{\hat\theta}$ can generated a plausible denoised image set $\small{\hat{\mathcal{X}}}$. Finally, the clean image can be estimated by averaging the images in $\small{\hat{\mathcal{X}}}$ using the MMSE estimation.}
    \label{Fig1}
\end{figure*}

\subsection{Electron Microscopy (EM) Denoising}
Conventional spatial filter based methods have been applied to EM, such as Bilateral filter~\cite{tomasi1998bilateral,pantelic2006discriminative}, Non-local Means~\cite{buades2005non,wei2010optimized}, BM3D\cite{dabov2007bm3d}, \textit{etc}.
Recent, deep learning based methods have been applied in EM imaging~\cite{ede2021deep,kalinin2022machine}.
Refs.~\cite{quan2019removing,wang2020noise2atom} proposed using Cycle-GAN for STEM images denoising without paired training images.
{Chong \textit{et al.}~\cite{chong2023m} utilized paired noisy images achieving real-world optical and
electron microscopy data denoising.}
Refs.~\cite{lin2021temimagenet}~\cite{mohan2022deep} proposed the simulation-based denoising (SBD) framework creating large simulated datasets to train CNNs for denoising in a supervised learning manner. 
Mohan \textit{et al.}~\cite{mohan2022deep} introduced simulated paired clean/noisy HREM images of the same substance for various imaging parameters. 
However, creating large simulated datasets can be a time-consuming process, and 
{its performance will degrade sharply due to the domain gap between simulated and real noisy images.}
Thus, it is imperative to propose a zero-shot image denoising method that is robust to ultra-low SNR to enhance HREM data quality effectively.



\section{Methodology}

\begin{figure}[!t]
    \centering
    \includegraphics[width=\columnwidth]{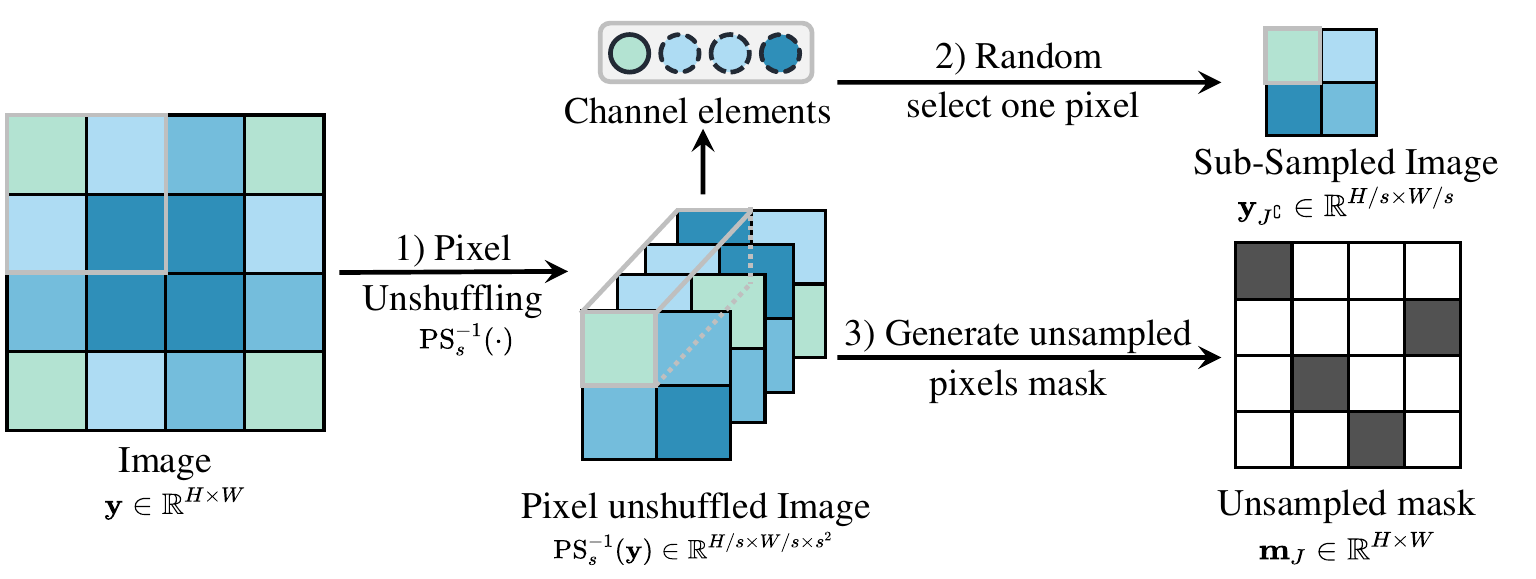}
    \caption{Illustration of the operations in the Random Sub-sampler with sampling stride $s$ to generate sub-sampled image $\mathbf{y}_{J^\complement}$ and unsampled mask $\mathbf{m}_J$.
    First, the input image is applied pixel unshuffling with a stride of $s$.
    At each location in the pixel unshuffled image, the Random Sub-sampler randomly selects $1$ element along the channel dimension to compose the sub-sampled image $\mathbf{y}_{J^\complement}$.
    {Meanwhile, the sampler sets the unsampled pixel mask $\mathbf{m}_J$ to 0 at the corresponding location if the element has been selected. Otherwise, it assigns a value of 1 (where black represents 0, and white represents 1). }
    The specific example in the figure demonstrates the sub-sampling process of the Random Sub-sampler with sampling stride $(s = 2)$ applied to an input image of size $4 \times 4$.}
    \label{Fig2}
\end{figure}

We introduce a novel self-supervised training framework called Noise2SR, which enables the training of a denoiser using paired noisy images with different resolutions obtained from an individual noisy observation. 
The effectiveness of our framework is supported by the theoretical proof of the $\mathcal{J}$-invariant property \cite{batson2019noise2self}.
Specifically, we introduce a novel random sub-sampling strategy to generate training image pairs of different resolutions. Subsequently, these pairs are utilized in conjunction with a super-resolution (SR) neural network. Such an approach effectively leverages the inherent relationship in signal content while minimizing noise correlation, thus significantly enhancing the efficiency of the denoiser training process.
We provide a comprehensive overview of the training and inference stages within the Noise2SR framework. 
The overall architecture of our proposed Noise2SR framework is illustrated in Fig. \ref{Fig1}.

\subsection{Related Theory Revisit}

\subsubsection{Noise2Noise} With the absence of lack of clean images for supervised learning, Noise2Noise (N2N) \cite{lehtinen2018noise2noise} proposed to train denoising network with paired noisy measurements of the same scene. 
Given paired noisy measurements $\{ \mathbf{y}_1, \mathbf{y}_2\} $ where $\{ \mathbf{y}_1 = \mathbf{x} + \mathbf{n}_1$ and $\mathbf{y}_1 = \mathbf{x} + \mathbf{n}_2 \}$, N2N demonstrates that learning the mapping between paired noisy measurements yields the same solutions as supervised learning with clean images statistically:
\begin{equation}
    \mathbb{E}_{\mathbf{x},\mathbf{y}_1,\mathbf{y}_2}\left \| f_\theta(\mathbf{y}_1 ) - \mathbf{y}_2 \right \|_2^2
=  \mathbb{E}_{\mathbf{x},\mathbf{y}_1,\mathbf{y}_2}\left \| f_\theta(\mathbf{y}_1 ) - \mathbf{x} \right \|_2^2 + \sigma^2,
\end{equation}
where $\sigma^2$ is a constant the variance of noise $\mathbf{n}$.

\subsubsection{\texorpdfstring{$\mathcal{J}$}\ -invariant Denoiser} With the assumption that noise is zero-mean and pixel-wise independent, Noise2Self \cite{batson2019noise2self} proved that denoising network with individual noisy measurements is possible if the network is $\mathcal{J}$-invariant. 

\noindent\textbf{Definition 1} \cite{batson2019noise2self} \textit{Let $\mathcal{J}$ be a partition of the dimensions $\{1,\dots,n\}$ and let $J \in \mathcal{J}$. A function $f: \mathbb{R}^n \to \mathbb{R}^n$ is $J$-invariant if $f(\mathbf{x})_J$ does not depend on the value of $\mathbf{x}_J$. It is $\mathcal{J}$-invariant if is J-invariant for each $J\in \mathcal{J}$. }

\subsection{Super-resolved Based Denoising Methods}
\label{secIII-B}
We propose training a denoiser using pairs of noisy images with different resolutions, generated from an individual image. These paired images consist of a sub-sampled noisy image $S(\mathbf{y})$ and its corresponding full-resolution noisy image $\mathbf{y}$.
Following the N2N strategy, the self-supervised loss can be generally stated as 
\begin{equation}
    \min_{\theta}\mathbb{E}\|f_\theta(S(\mathbf{y}))-\mathbf{y}\|^2.
\end{equation}
However, directly minimizing the loss function above may result in the network learning an identity mapping for the noisy input pixels.
Thus, we proposed to use unsampled noisy pixels for network supervision.
Under this supervision, we can theoretically prove that using noisy images for network training is equivalent to using clean signals for supervision.

Based on the $\mathcal{J}$-invariant theory, the sub-sampling operation partitions the image with $n$ pixels into two disjoint sets denoted as $J$ and $J^\complement$, where $|J| + |J^\complement| = n$. Consequently, the sub-sampled noisy image $S(\mathbf{y})$ can be represented as $\mathbf{y}_{J^\complement}$, while the unsampled noisy pixels $\mathbf{y} \setminus S(\mathbf{y}) $ are denoted as $\mathbf{y}_J$. When training a network using paired noisy images with different resolutions, the resulting network naturally becomes a $\mathcal{J}$-invariant function. This is because the output $f_\theta(\mathbf{y}_{J^\complement})_J$ does not rely on the specific values of $\mathbf{y}_J$.

\noindent {\bf{Theorem 1.}}
\textit{Let $\mathbf{y} = \mathbf{x} + \mathbf{n}$ be an image corrupted by zero-mean noise $\mathbf{n}$ with variance $\sigma^2$. $\mathbf{y}_{J^\complement}$ is the sub-sampled noisy image $\mathbf{y}$ and $\mathbf{y}_{J}$ is the unsampled noisy pixels of original image $\mathbf{y}$. Suppose the noise $\mathbf{n}$ is independent of the clean image $\mathbf{x}$ and the noise $\mathbf{n}$ is pixel-wise independent. Then it holds that}:
\begin{equation}
\mathbb{E}_{\mathbf{x},\mathbf{y}}\left \| f_{\theta} (\mathbf{y}_{J^\complement})_{J}- \mathbf{y}_{J}\right \|_2^2 =
 \mathbb{E}_{\mathbf{x},\mathbf{y}}\left \| f_{\theta} (\mathbf{y}_{J^\complement})_{J}- \mathbf{x}_{J}\right \|_2^2 + \sigma^2.
\end{equation}
The proof is given below:

\noindent {\bf{Proof :}}
\begin{equation}
\begin{aligned}
    \mathbb{E}_{\mathbf{x},\mathbf{y}}\left \| f_\theta(\mathbf{y}_{J^\complement} )_{J} - \mathbf{y}_{J} \right \|_2^2 
 & =   \mathbb{E}_{\mathbf{x},\mathbf{y}}\left \| f_\theta(\mathbf{y}_{J^\complement} )_{J} - \mathbf{x}_{J} - \mathbf{n}_{J} \right \|_2^2 \\
  & =   \mathbb{E}_{\mathbf{x},\mathbf{y}} \left \| f_\theta(\mathbf{y}_{J^\complement})_{J} - \mathbf{x}_{J} \right \|_2^2 + \sigma^2 \\
&\quad -  2 \mathbb{E}_{\mathbf{x},\mathbf{y}}\langle f_{\theta}(\mathbf{y}_{J^\complement})_{J} - \mathbf{x}_{J} , \mathbf{n}_{J}   \rangle.
\end{aligned}
\end{equation}
Due to the independence between $\mathbf{y}_{J^\complement}$ and $\mathbf{x}_{J}$ to $\mathbf{n}_{J}$, there holds:
\begin{equation}
\begin{aligned}
        \mathbb{E}_{\mathbf{x},\mathbf{y}} \langle f_\theta(\mathbf{y}_{J^\complement})_{J} - \mathbf{x}_{J}, \mathbf{n}_{J}\rangle 
 = \mathbb{E}_{\mathbf{x},\mathbf{y}} \left [ f_\theta(\mathbf{y}_{J^\complement})_{J} - \mathbf{x}_{J} \right ]  \mathbb{E}_{\mathbf{x},\mathbf{y}}
        [\mathbf{n}_{J}].
\end{aligned}
\end{equation}
Since the noise is zero-mean $\mathbb{E}_{\mathbf{x},\mathbf{y}}(\mathbf{n}_{J}) = 0$, we have:
\begin{equation}
    \mathbb{E}_{\mathbf{x},\mathbf{y}}\left \| f_{\theta} (\mathbf{y}_{J^\complement})_{J}- \mathbf{y}_{J}\right \|_2^2 =
 \mathbb{E}_{\mathbf{x},\mathbf{y}}\left \| f_{\theta} (\mathbf{y}_{J^\complement})_{J}- \mathbf{x}_{J}\right \|_2^2 + \sigma^2.
\end{equation}
Since $\sigma^2$ is constant, Theorem 1 states that optimizing the self-supervised loss function over the proposed training scheme yields the same solutions as the supervised loss function.

\subsection{Generating Sub-sampled Image Randomly}
Self-supervised image denoising methods typically assume that noise is signal-independent and spatially uncorrelated. 
Therefore, it is crucial to sub-sample a noisy image to maintain the denoising performance when training with paired noisy images of different resolutions.
Recently, pixel-shuffling downsampling (PD) \cite{zhou2020awgn,lee2022ap} has emerged as an effective technique for breaking the spatial correlation of real-world noise. 
Motivated by the success of PD, we introduce a Random Sub-sampler for generating sub-sampled noisy images. 
The Random Sub-sampler follows a similar sub-sampling strategy as PD but introduces additional randomness into the sub-sampling operation.
The randomness serves as a data augmentation strategy, which helps to prevent training overfitting.

The process of using Random Sub-sampler with sampling stride $s$ to generate a sub-sampled noisy image $\mathbf{y}_{J^\complement} \in \mathbb{R}^{H/s \times W/s}$ from an image $\mathbf{y}\in \mathbb{R}^{H\times W}$  is summarized as follows:

\begin{enumerate}
    \item Perform an inverse pixel shuffling (PS) \cite{shi2016real} operation on image  $\mathbf{y} \in \mathbb{R}^{H\times W}$  with a stride of $s$, resulting in:    $\text{PS}_s^{-1}(\mathbf{y}) \in \mathbb{R}^{H/s\times W/s \times s^2}$;
    \item For $(i,j)$-th location of sub-sampled image $\mathbf{y}_{J^\complement}$, randomly select one elements from  ${\text{PS}_{s}^{-1}(\mathbf{y})}_{ij}$;
    \item Generate a binary matrix mask $\mathbf{m}_J \in \mathbb{R}^{H\times W}$ to select unsampled pixels in the original image $\mathbf{y}$. 
    The mask $\mathbf{m}_J$ is set to 0 at locations where an element from $\mathbf{y}$ is selected, and 1 otherwise.
\end{enumerate}
Following this process, the sub-sampled noisy image $\mathbf{y}_{J^\complement}$ with dimensions of $H/s \times W/s$ is obtained.
The binary matrix mask $\mathbf{m}_J$ is also generated to identify the unsampled pixels in $\mathbf{y}$.
Fig.~\ref{Fig2} illustrates the workflow of the Random Sub-sampler, demonstrating the generation of a sub-sampled image from an input image of size $4\times 4$ with a stride of $s = 2$.

\begin{algorithm}[t]
\caption{Zero-shot learning for Noise2SR}\label{alg:alg1}
\begin{algorithmic}[1]
\REQUIRE An individual noisy image $\mathbf{y}$; denoising network $f_\theta$; Random Sub-sampler. 
\ENSURE Well-trained denoising network $f_{\hat{\theta}}$.
\\

\WHILE{\textit{not converged}} 
\STATE Generate a random sub-sampled noisy image $\mathbf{y}_{J^\complement}$, and a binary mask $\mathbf{m}_J$ from a Random Sub-sampler;
\STATE For a sub-sampled noisy image $\mathbf{y}_{J^\complement}$, derive the denoised image $f_{\theta}(\mathbf{y}_{J^\complement})$;
\STATE Select the unsampled pixels in noisy image $\mathbf{y}$ and corresponding pixels in $f_{\theta}(\mathbf{y}_{J^\complement})$:\\
\hspace{0.2cm} $\mathbf{y}_J = \mathbf{y} \odot \mathbf{m}_J$, 
\hspace{0.1cm} $f_{\theta}(\mathbf{y}_{J^\complement})_J = f_{\theta}(\mathbf{y}_{J^\complement})\odot \mathbf{m}_J$;
\STATE Calculate $\mathcal{L} = \| f_{\theta}(\mathbf{y}_{J^\complement})_J -\mathbf{y}_J \|^2$;
\STATE Update the denoising network $f_\theta$ by minimizing the objective
$\mathcal{L}$.

\ENDWHILE
\end{algorithmic}
\label{alg1}
\end{algorithm}

In section~\ref{secIV-F}, we evaluate the effectiveness of randomness in sub-sampler on the performance.
The stride $s$ determines the sub-sampling interval, which helps to break the spatial correlation of noise. It also controls the sampling ratio of noisy pixels and affects the relative receptive field of the network, both of which influence the training process of Noise2SR.
In Section~\ref{secIV-E},  we conduct a comprehensive study to evaluate the influence of the sampling stride $s$ of the Random Sub-sampler. 
Recent studies \cite{huang2022neighbor2neighbor,pan2023random} adopt similar sub-sampling strategies. However, our Random Sub-sampler generates paired noisy images at different resolutions while their sub-sampled images are at the same resolution.

\subsection{Optimizing Network}
Given a noisy image $\mathbf{y} \in \mathbb{R}^{H\times W}$, We can parameterize the Noise2SR as CNN-based denoising and SR function $f_{\theta}: \mathbb{R}^{|J^\complement|} \to \mathbb{R}^{H\times W}$, where the parameters $\theta$ are weights of the network, $|J^\complement|$ is number of sub-sampled noisy image pixels.
Noise2SR takes the sub-sampled noisy image $\mathbf{y}_{J^\complement}$ as input and outputs a prediction of denoised image $\hat{\mathbf{x}} = f_\theta(\mathbf{y}_{J^\complement}) \in \mathbb{R}^{H\times W}$.

Based on the proof provided in Section \ref{secIII-B}, we optimize the network by minimizing the loss function that compares the unsampled pixels of the original noisy image denotes as $\mathbf{y}_J$, with corresponding pixels in the network output, denotes as $f_\theta(\mathbf{y}_{J^\complement})_J$.
To facilitate the selection of these pixels, we can utilize a binary mask $\mathbf{m}_J$. Specifically, the unsampled pixels of the original image can be obtained by element-wise multiplication with the mask, \textit{i.e.}, $\mathbf{y}_J = \mathbf{y} \odot \mathbf{m}_J$. Similarly, the corresponding pixels in the network output can be obtained as $f_\theta(\mathbf{y}_{J^\complement})_J = f_\theta(\mathbf{y}_{J^\complement}) \odot \mathbf{m}_J$. Here, the symbol $\odot$ represents the Hadamard product.
Thus, the Noise2SR network can learn a denoising and SR function by minimizing the self-supervised loss:
\begin{equation}
\underset{\theta}{\arg\min}\sum_{J\in\mathcal{J}}\mathcal{L}(f_\theta(\mathbf{y}_{J^\complement})_J, \mathbf{y}_{J}),
\end{equation}
where $\mathcal{J}$ represents the set of partitions used during the training stage. 

\subsection{Clean Image Restoration with MMSE Estimation}
The pipeline of clean image reconstruction using the well-trained N2SR model is shown in Fig.~\ref{Fig1}.B. 
For a random sub-sampled noisy image $\mathbf{y}_{J^\complement}$, well-trained Noise2SR can generated a plausible denoised result $\hat{\mathbf{x}} = f_{\hat{\theta}}(\mathbf{y}_{J^\complement})$.
Since the sub-sampled image is generated from the original noisy image randomly, the clean image can be expressed as:
\begin{equation}
    \mathbf{x} = \mathbb{E}[f_{\hat{\theta}}(\mathbf{y}_{J^\complement})] 
= \sum_{J\in\mathcal{J}} f_{\hat{\theta}}(\mathbf{y}_{J^\complement}) p(\mathbf{x} | \mathbf{y}_{J^\complement}),
\end{equation}
where $\mathcal{J}$ is the possible Random Sub-sampler partition set.
Since the randomness of sub-sampling operation , $|\mathcal{J}|$ is numerous and $p(\mathbf{x}|\mathbf{y}_{J^\complement})$ is undetermined, making it challenging to precisely compute $\mathbb{E}(\hat{\mathbf{x}})$.

Thus, we propose to use MMSE estimation to approximate $\mathbb{E}(\hat{{\mathbf{x}}})$ of the clean signal.
Given a set of sub-sampled noisy images $\mathbf{y}_{J^\complement}$,  we can approximate the MMSE estimate of the clean image by averaging all the plausible denoised results $f_{\hat \theta} (\mathbf{y}_{J^\complement})$.
Specifically, we first randomly sub-sampled the noisy image $M$ times to  obtain a sub-sampled noisy image set  $ \mathcal{Y} = \left \{\mathbf{y}_{J_1^\complement}, \dots, \mathbf{y}_{J_M^\complement} \right \}$. 
For each subsampled noisy image $\mathbf{y}_{J_m^\complement}$, the well-trained Noise2SR takes it as input and generates the corresponding denoising and SR result $f_{\hat \theta} \left (\mathbf{y}_{J_m^\complement}\right )$. 
Finally, the approximate MMSE estimation of the clean image can be computed below: 
\begin{equation}
    \mathbf{x} \approx \frac{1}{M}\sum_{m=1}^{M} f_{\hat{\theta}} \left ( \mathbf{y}_{J_m^\complement} \right),
\end{equation}
where $J_m^\complement$ denotes the sub-sampling partition at $m$ time.

\begin{figure}[!t]
    \centering
    \includegraphics[width=\columnwidth]{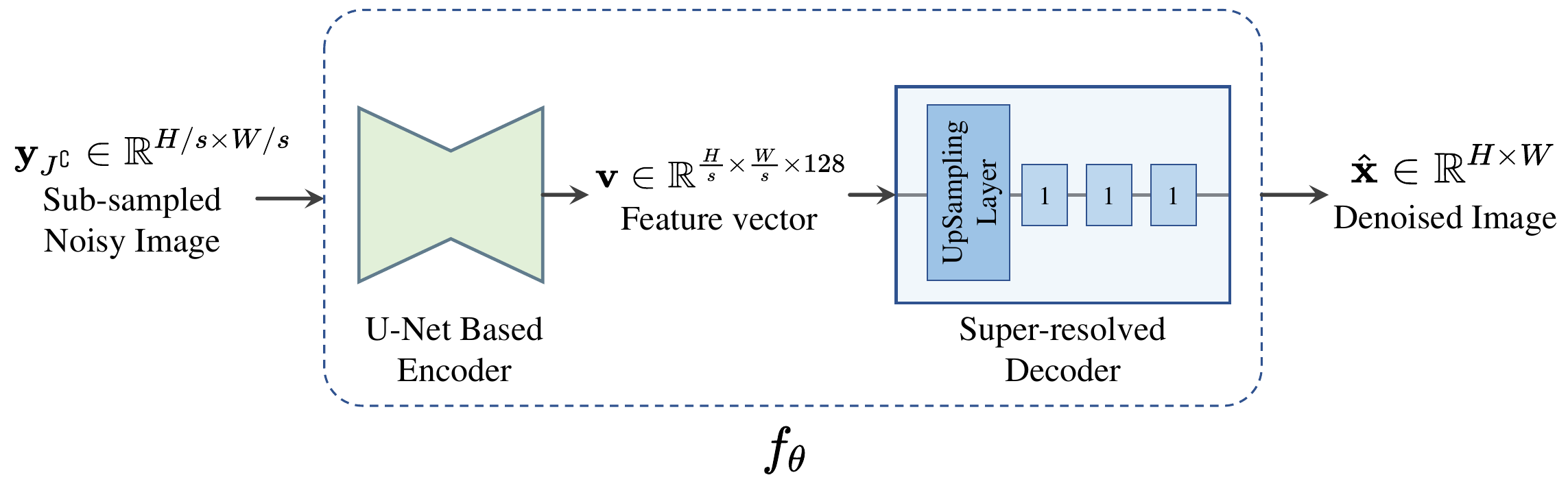}
    \caption{The architecture of Noise2SR used for parameterizing the super-resolved denoising function $f_\theta$, which consists of the U-Net based encoder and thesuper-resolved decoder.}
    \label{Fig3}
\end{figure}

\subsection{Network Architecture}
As shown in Fig.~\ref{Fig3}, the network of Noise2SR $f_{\theta}$ consists of a U-Net based encoder and a Super-resolved decoder. 
\subsubsection{U-Net based Encoder}
We adopt the same U-Net architecture as encoder \cite{lehtinen2018noise2noise} while modifying the last convolution block to generate a feature vector $\mathbf{v}$ with 128 channels for the Super-resolved decoder.
\subsubsection{Super-resolved Decoder}
The decoder takes the feature vector as input to generate the denoised image $\hat{\mathbf{x}}$. It consists of an upsampling layer followed by three convolutional layers, each with a 1$\times$1 kernel size.
In this work, we employ pixel-shuffling \cite{shi2016real} as the super-resolution strategy for the upsampling layer.
In Section \ref{secIV-E}, we conduct an ablation study to evaluate the impact of different super-resolution strategies on denoising performance.

\section{Experiments and Results}
\begin{figure*}[!t]
    \centering
    \includegraphics[width=\textwidth]{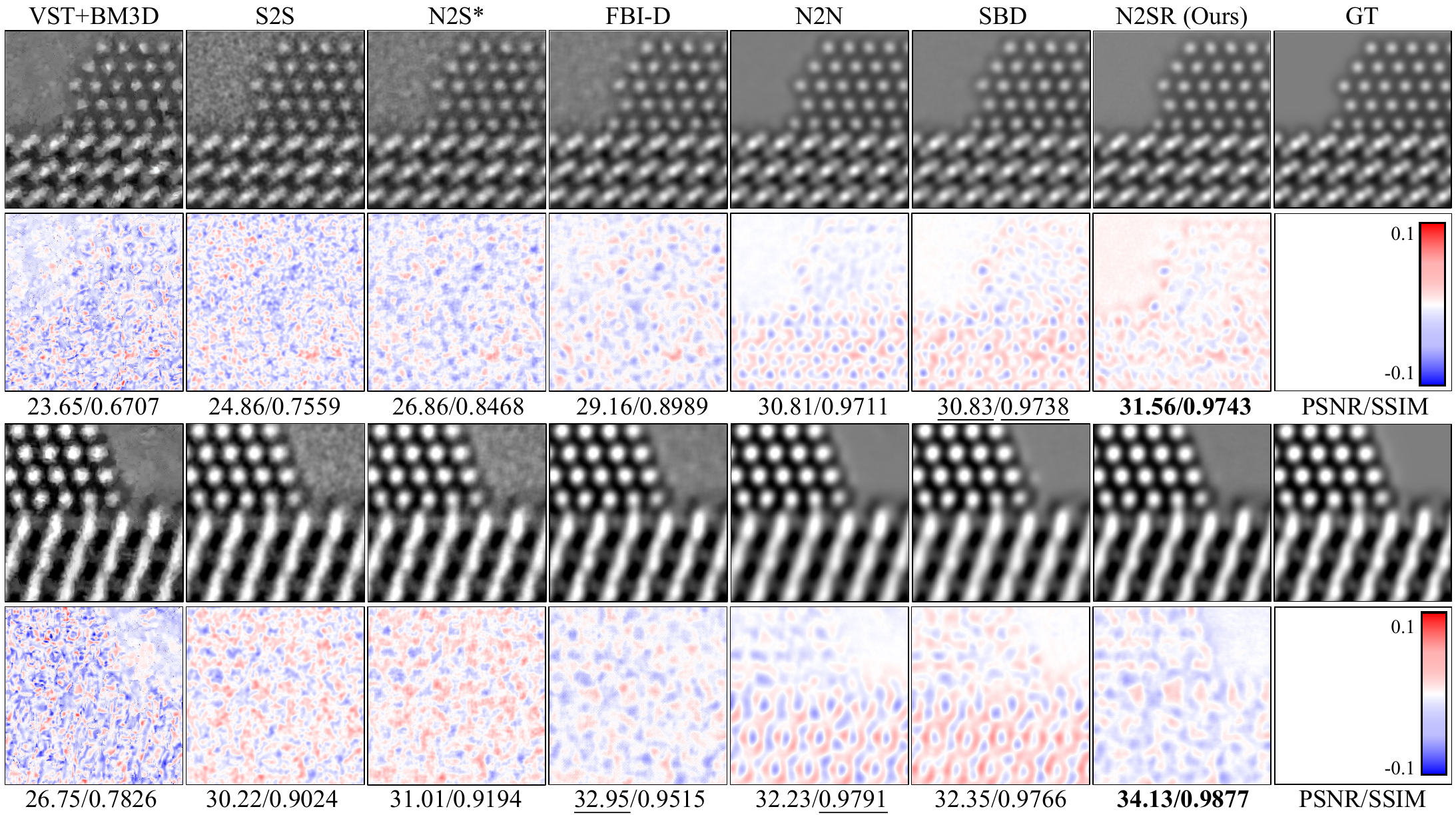}
    \caption{Comparison of Noise2SR and other image denoising methods on simulated \ce{Pe/CeO2} catalyst corrupted with Poisson-Gaussian noise $(a = 0.05, b = 0.02)$. 
    The second and fourth rows display the corresponding error maps of the denoised results.
    * indicates that the dataset-based self-supervised denoising method was performed in a zero-shot learning manner.}
    \label{res_f1}
\end{figure*}

\begin{figure}[!t]
    \centering
    \includegraphics[width=\columnwidth]{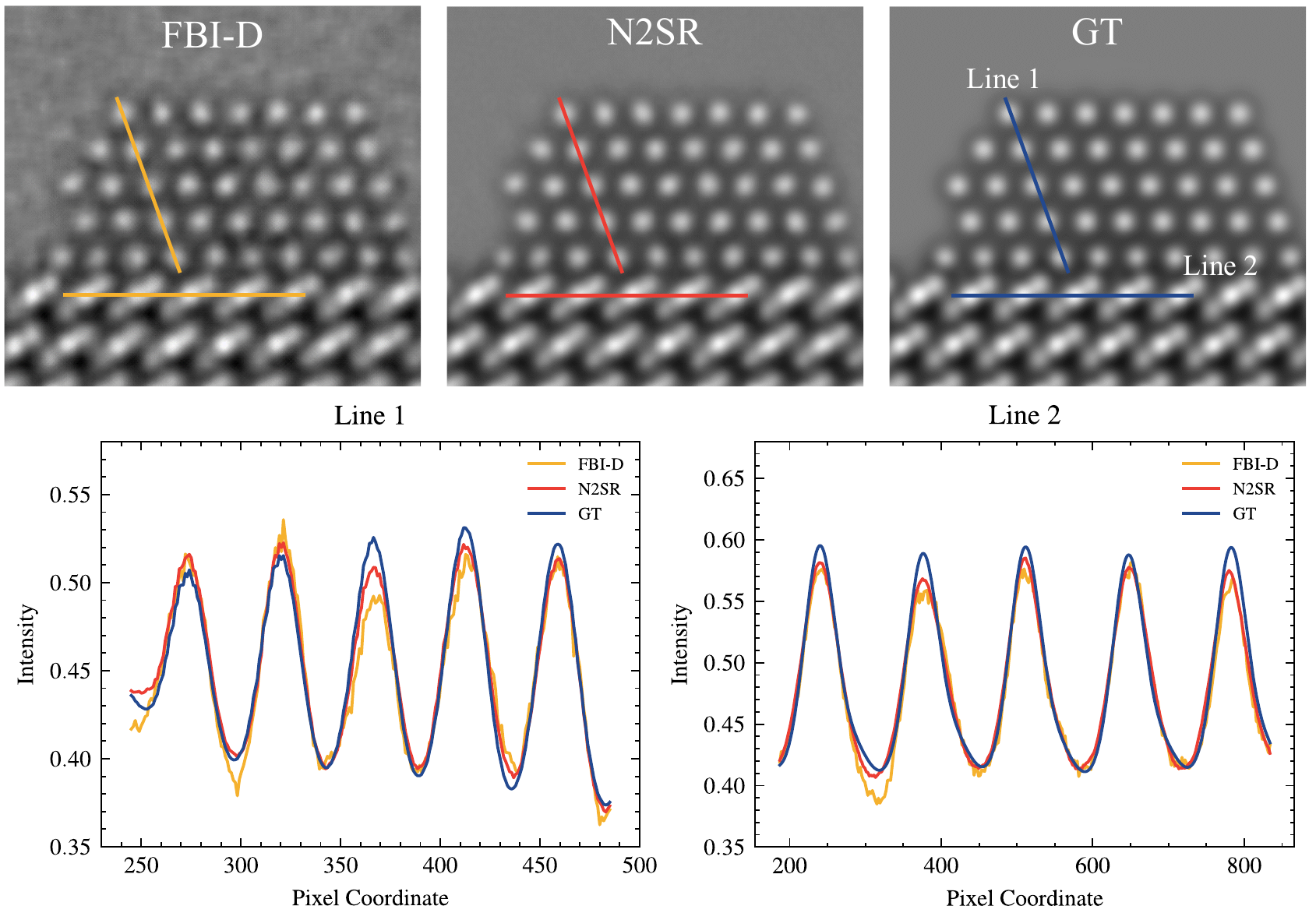}
    \caption{Comparison of intensity profiles on the surface atomic columns is conducted for the denoised results obtained using FBI-Denoiser (FBI-D) and Noise2SR (N2SR) on simulated \ce{Pe/CeO2} catalyst corrupted with Poisson-Gaussian noise $(a = 0.05, b = 0.02)$, alongside the corresponding ground truth data.}
    \label{res_profile}
\end{figure}

In this section, we aim to demonstrate the feasibility and effectiveness of the proposed Noise2SR framework.
First, we evaluate the performance of Noise2SR compared with eight other denoising methods on simulated and real noise removal in HREM data.
Then, we conduct comprehensive ablation studies to analyze the impact of key components, including the Random Sub-sampler, MMSE estimation of the clean signal, and the architecture of the encoder and decoder in the Noise2SR network.
Finally, we demonstrate some failure cases to demonstrate the extreme performance limits of Noise2SR
and discuss the potential approaches for improvement. 

\subsection{Compared Methods and Metrics}
\subsubsection{Compared Methods}
For comprehensive comparisons, we compared the proposed Noise2SR with 8 methods which are divided into four groups.
1) non-learning methods: Adaptive Wiener filtering \cite{lim1990awf} and BM3D\cite{dabov2007bm3d}; 
2) Zero-shot Self-supervised Learning (ZS-SSL) methods: 
Self2Self (S2S) \cite{quan2020self2self}, FBI-Denoiser (FBI-D) \cite{byun2021fbi} and ZS-Noise2Noise~\cite{mansour2023zero};
3) Dataset-based Self-supervised Learning (DS-SSL) methods:
Noise2Self (N2S) \cite{batson2019noise2self}, Neighbor2Neighbor (NB2NB) \cite{huang2022neighbor2neighbor} and Noise2Noise (N2N) \cite{lehtinen2018noise2noise};
4) Supervised Deep learning methods: Simulated-based denoising (SBD) \cite{mohan2022deep}. 
\textit{ It should be noted that N2S and NB2NB can also be trained in zero-shot learning manner, and we denote the corresponding extensions of these methods as N2S* and NB2NB*, respectively.}

\subsubsection{Implementation Details}
{For the training of the Noise2SR model, each iteration begins by randomly cropping a patch image of size 128 $\times$ 128 from the original image. }
Subsequently, we utilize the Random Sub-sampler with sampling stride $(s = 2)$ to generate a sub-sampled noisy image, which serves as the input for the network.
To optimize the model, we employ the Adam optimizer \cite{kingma2014adam} with the following hyper-parameters: $\beta_1 = 0.9$, $\beta_2 = 0.999$, and $\epsilon = 10^{-8}$. 
The initial learning rate is set to $10^{-4}$.
We set the batch size of 12 and the training process consists of 1500 epochs.
During the inference stage, we sub-sampled 50 times to generate an approximate MMSE denoised result.

For the compared DL-based methods, we adhere to the same network architecture as described in their respective original papers. 
Specifically, for the SBD method, we employ the U-Net architecture \cite{lehtinen2018noise2noise}, also referred to as small-Unet in the original paper.
All experiments were conducted on a server equipped with Python 3.7.3, PyTorch 1.3, and NVIDIA TITAN GPUs.

\begin{table*}[!t]
\centering
\caption{Quantitative results (PSNR/SSIM) of compared denoising methods and Noise2SR on simulated \ce{Pt/CeO2} TEM data corrupted with Poisson-Gaussian noise. * indicates that the dataset-based self-supervised denoising methods are performed in a zero-shot learning manner. The best performance among non-learning and zero-shot learning methods is highlighted in \textbf{bold}, while the second-best performance is \underline{underlined}. }
\label{tab1}
\renewcommand\arraystretch{1}
\resizebox{\textwidth}{!}{
\begin{tabular}{clcccccc} 
\toprule
\multicolumn{2}{c}{\textbf{Noise Parameters}}  & \multicolumn{2}{c}{$a = 0.1, b = 0.02$}             & \multicolumn{2}{c}{$a = 0.05,  b = 0.02$}            & \multicolumn{2}{c}{$a = 0.02, b = 0.02$}             \\ 
\cmidrule(lr){3-4}\cmidrule(r){5-6}\cmidrule(r){7-8}
\textbf{Category}                                                                                 & \multicolumn{1}{c}{\textbf{Method}}          & PSNR                    & SSIM                     & PSNR                    & SSIM                     & PSNR                    & SSIM                      \\ 
\midrule
\text{Noisy} &  & 4.57$\pm$2.67          & 0.0120$\pm$0.01          & 7.54$\pm$2.67         & 0.0232$\pm$0.01          & 11.40$\pm$2.67          & 0.0526$\pm$0.02 \\ 
\midrule
\multirow{2}{*}{\begin{tabular}[c]{@{}c@{}}\text{Non-} \\\text{Learning} \end{tabular}}                                                       & Adaptive Wiener~\cite{lim1990awf}                      & 20.70$\pm$1.90          & 0.6093$\pm$0.08          & 22.96$\pm$1.97          & 0.7398$\pm$0.07          & 25.66$\pm$1.95          & 0.8656$\pm$0.03           \\
                                                                                         & VST+BM3D~\cite{dabov2007bm3d}                             & 22.30$\pm$2.34          & 0.5429$\pm$0.10          & 25.33$\pm$2.33          & 0.7266$\pm$0.08          & 29.97$\pm$2.33          & 0.8659$\pm$0.05           \\ 
\midrule
\multirow{7}{*}{\begin{tabular}[c]{@{}c@{}}\text{ZS-SSL} \end{tabular}}            & Neighbour2Neighbour*~\cite{huang2022neighbor2neighbor}                               & 24.56$\pm$2.48          & 0.6983$\pm$0.10          & 26.32$\pm$2.61          & 0.7451$\pm$0.12          & 28.44$\pm$2.87          & 0.7915$\pm$0.09           \\
                                                                                         & Noise2Self*~\cite{batson2019noise2self}                                 & 28.51$\pm$2.99  & 0.8817$\pm$0.06  & 29.76$\pm$2.36          & 0.9056$\pm$0.04          & 31.68$\pm$2.69          & 0.9207$\pm$0.06           \\
                                                                                         & Self2Self~\cite{quan2020self2self}                                  & 25.36$\pm$2.79          & 0.7423 $\pm$0.13         & 27.77$\pm$2.93          & 0.8353$\pm$0.07          & 31.05$\pm$2.50          & 0.9076$\pm$0.04           \\
                                                                                         & FBI-D~\cite{byun2021fbi}                              & 28.17$\pm$2.73          & 0.7916$\pm$0.09          & \uline{31.23$\pm$2.74}  & 0.9231$\pm$0.04  & 34.23$\pm$2.74  & 0.9541$\pm$0.04   \\
                                                                                         & ZS-N2N~\cite{mansour2023zero}                              & 21.95$\pm$2.00          & 0.5776$\pm$0.07          & 24.34$\pm$2.31   & 0.6736$\pm$0.08   & 27.84$\pm$2.62  &  0.7910$\pm$0.07  \\
                                                                                         & Noise2SR (w/o MMSE) (Ours)                          & \underline{28.84$\pm$2.95} & \underline{0.9296$\pm$0.05} & 30.16$\pm$3.08 & \uline{0.9532$\pm$0.04} & \uline{34.57$\pm$2.46} & \uline{0.9788$\pm$0.01}  \\
                                                                                         & Noise2SR (w/ MMSE) (Ours)                          & \textbf{31.68$\pm$2.68} & \textbf{0.9656$\pm$0.02} & \textbf{33.66$\pm$2.68} & \textbf{0.9772$\pm$0.01} & \textbf{36.35$\pm$2.46} & \textbf{0.9873$\pm$0.01}  \\ 
\midrule
\multirow{3}{*}{\begin{tabular}[c]{@{}c@{}}\text{DS-SSL} \end{tabular}} & Noise2Self\cite{batson2019noise2self}                                  & 31.62$\pm$2.77          & 0.9642$\pm$0.02          & 33.50$\pm$2.58          & 0.9734$\pm$0.01          & 35.68$\pm$2.91          & 0.9848$\pm$0.01           \\
                                                                                         & Neighbour2Neighbour~\cite{huang2022neighbor2neighbor}                                & 28.80$\pm$3.10          & 0.9305$\pm$0.03          & 30.75$\pm$2.37          & 0.9530$\pm$0.02          & 33.21$\pm$2.72          & 0.9674$\pm$0.01           \\
                                                                                         & Noise2Noise~\cite{lehtinen2018noise2noise}                                  & 32.73$\pm$2.97          & 0.9790$\pm$0.01          & 34.11$\pm$2.88          & 0.9837$\pm$0.01          & 36.30$\pm$2.78          & 0.9888$\pm$0.01           \\ 
\midrule
\text{Supervised}                                                                              & SBD~\cite{mohan2022deep}                                  & 33.12$\pm$2.79          & 0.9799$\pm$0.01          & 34.25$\pm$3.09          & 0.9838$\pm$0.01          & 36.52$\pm$2.90          & 0.9891$\pm$0.01           \\
\bottomrule
\end{tabular}
}
\end{table*}

\subsubsection{Evaluation Metrics}
We calculate Peak-Signal-to-Noise Ratio (PSNR) and Structured Similarity Index Measure (SSIM)~\cite{wang2004image} to measure the performance of compared methods quantitatively. 
PSNR is defined based on pixel-by-pixel distance and SSIM measures structural similarity using the mean and variance of images.

\subsection{Datasets}
\noindent\textbf{Simulated {Noisy} TEM Datasets}

\subsubsection{Simulated TEM Datasets}
Based on the simulated TEM dataset
\footnote{\url{https://github.com/sreyas-mohan/electron-microscopy-denoising}} released by Mohan \textit{et al}. \cite{mohan2022deep} which consists of approximate 18000 simulated images of \ce{Pt/CeO2} catalyst with various combinations of imaging parameters, we select 1000 images as training set and 5 images for the test set. 
\textit{The training set is used to prepare dataset-based training methods and evaluate their performance on the test set. On the other hand, zero-shot learning models train a model for each image in the test set.}

\subsubsection{Noise simulation}
HREM images are affected by mixed Poisson-Gaussian noise~\cite{boyat2015review,sarder2006deconvolution,meiniel2018denoising}, which combines the effects of dark noise and photon noise (Poisson noise) and readout noise (Gaussian noise). 
The Poisson noise $\mathbf{n}_{p}$ is dependent on the signal intensity, while the Gaussian noise $\mathbf{n}_{g}$ originates from imperfections in the output amplifier during charge-to-voltage conversion. We adopt the Poisson-Gaussian Model formulation described in~\cite{bahler2022pogain}.
\begin{equation}
\begin{aligned}
& \mathbf{y} = \mathbf{x} + \mathbf{n}_{p}(\mathbf{x}) + \mathbf{n}_{g},\\
& (\mathbf{x} + \mathbf{n}_{p}(\mathbf{x}))\sim {a}\mathcal{P}\left (\frac{1}{a}\mathbf{x} \right ),\\
& \mathbf{n}_{g} \sim \mathcal{N}(0,b^2).
\end{aligned}
\end{equation}
The expected value and the variance of the noisy measurement $\mathbf{x}$ is:
\begin{equation}
    \mathbb{E}[\mathbf{y}] = \mathbf{x}, \quad \text{Var}[\mathbf{x}] = a\mathbf{x} + b^2.
\end{equation}
Thus, the noisy image corrupted with Poisson-Gaussian noise can be modeled using characterized  parameters $(a,b)$ as
\begin{equation}
    \mathbf{y} = \mathbf{x} + \mathbf{n}, \quad \mathbf{n} \sim \mathcal{N}(0,a\mathbf{x} + b^2).
\end{equation}
Here, $a$ represents the noise level that is dependent on the signal, while $b$ represents the noise level that is independent of the signal.

To comprehensively evaluate the denoising performance of the compared methods on different levels of noise, we apply three different levels of Poisson-Gaussian noise to the TEM dataset. The noise parameters used are summarized below: $(a=0.1, b=0.02), (a=0.05, b=0.02),$ and $(a=0.01, b=0.02)$.

\begin{figure*}[!t]
    \centering
    \includegraphics[width=\textwidth]{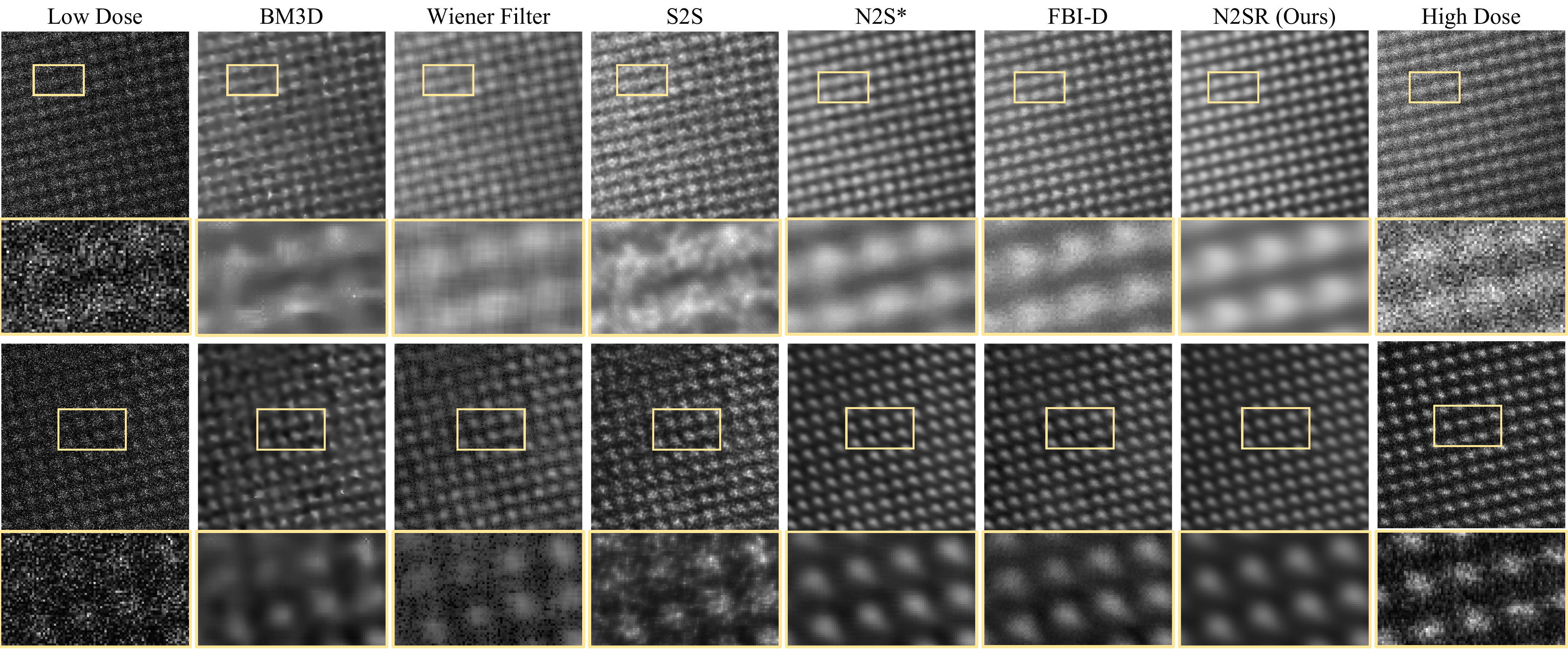}
    \caption{Comparison of Noise2SR with other image denoising methods on real low-dose Te STEM images. 
    The last column displays the corresponding Te imaging under high-dose conditions for reference.
    * indicates that the dataset-based self-supervised denoising method was performed in a zero-shot learning manner.}
    \label{res_Te}
\end{figure*}

\noindent\textbf{Real STEM Data} Sample preparation of Te crystals, ZSM-5 (MFI framework) and MOR zeolite 
(\url{http://asia.iza-structure.org/IZA-SC/ftc_table.php}) were all followed in the same way. 
The crystals were first crushed with mortar and pestle, and then the powders were dispersed in ethanol under ultrasonication. Few drops of the suspension were placed onto holey carbon copper grids to be further checked by TEM. 
The real annular dark field scanning transmission electron microscopy (ADF-STEM) images were all collected with a GrandARM 300F (JEOL Ltd.) transmission electron microscopy operated at 300 kV. 
The crystals were first tilted to specific zone axis under TEM mode with the selected area electron diffraction (SAED) patterns and then switched to STEM mode to record the experimental ADF-STEM images.

\subsection{Comparisons on Simulated Noisy TEM Datasets}
Table~\ref{tab1} shows the quantitative results. Compared with other zero-shot self-supervised learning denoising methods, Noise2SR achieves the best denoising performance. Compared with dataset-based self-supervised denoising methods and supervised denoising, Noise2SR achieves comparable performance.

Fig.~\ref{res_f1} shows the qualitative results on two test samples of the simulated noisy \ce{Pt/CeO2} catalyst TEM images. Compared with zero-shot denoising methods, Noise2SR precisely restores the nanoparticle structures and achieves a cleaner background.  Moreover, Noise2SR exhibits fewer additional structural patterns in the error maps compared to dataset-based self-supervised denoising methods and supervised denoising methods.
In Fig.~\ref{res_profile}, we compare the intensity profiles on the surface atomic columns for the denoised data obtained from FBI-D and Noise2SR, along with the ground truth. 
The intensity profiles of both denoised data show a similar overall trend to the ground truth. 
Compared to FBI-D, Noise2SR exhibits a closer resemblance to the ground truth in terms of both peaks and troughs, with fewer fluctuations.

\subsection{Comparisons on Real STEM Data}
\subsubsection{Low-Dose Te Crystal STEM Data}
In Fig.~\ref{res_Te}, we present the qualitative results obtained from denoising real low-dose Te crystal STEM data using different methods. 
The corresponding high-dose Te crystals STEM data is also included for reference.
Two different areas of the low-dose and high-dose STEM images of one same Te crystal were shown in Fig.~\ref{res_Te}, where bright contrast that corresponds to each Te atom can be observed while the contrast is less clear in low-dose image than high-dose image because of the low signal-to-noise ratio (SNR).
After denoising of low-dose STEM images using different methods, it can be observed that conventional learning methods (\textit{e.g.}, BM3D, Wiener Filter) tend to produce over-smoothed results, making it difficult to distinguish the contours of the atoms.
On the other hand, DL-based denoising methods exhibit good atom contrast so that each Te atom can be clearly resolved. 
Notably, compared to other zero-shot denoising methods, the denoised results of Noise2SR exhibit reduced noise and clearer atom contours.
\begin{figure*}[!t]
    \centering
    \includegraphics[width=\textwidth]{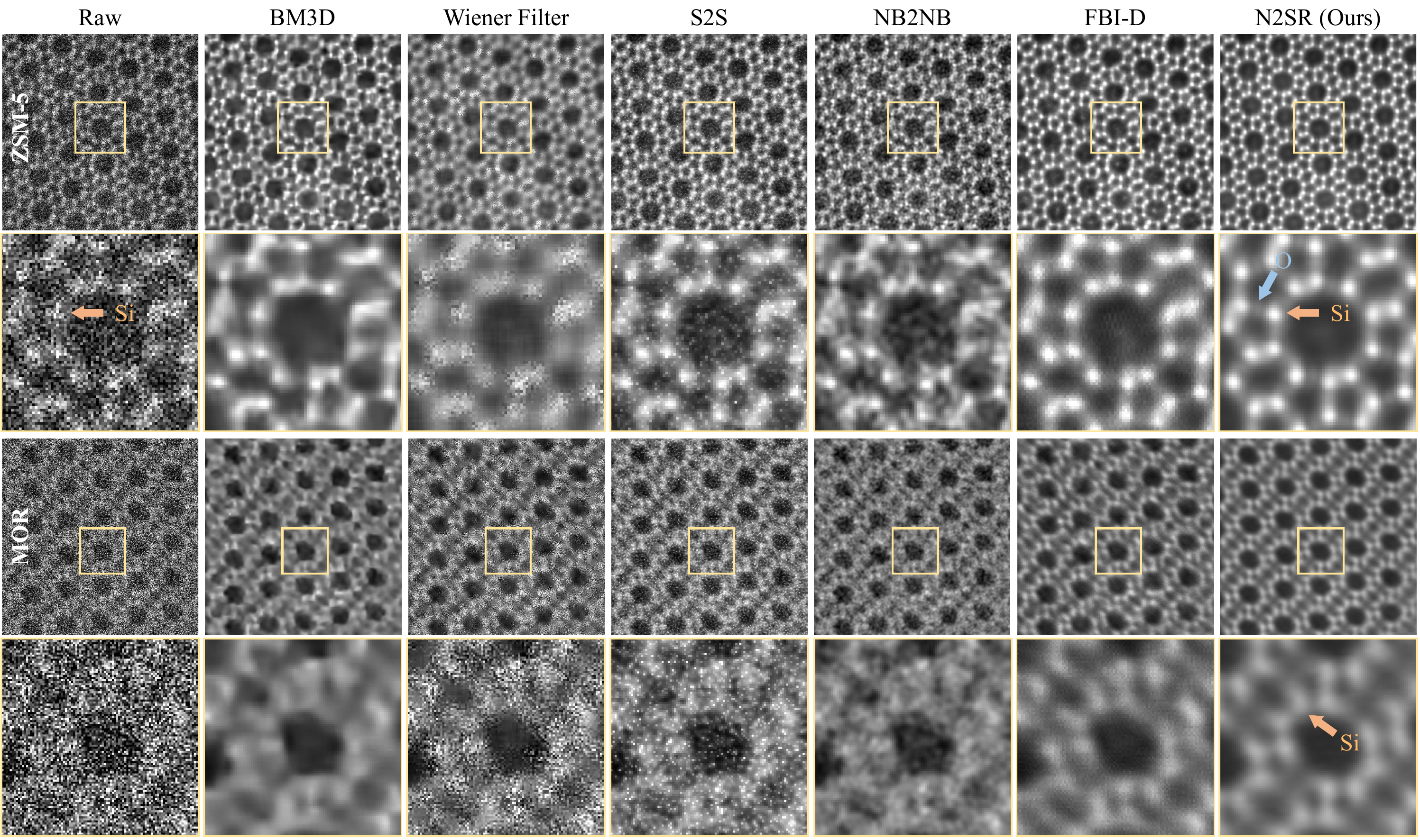}
    \caption{Comparison of other image denoising methods with Noise2SR on real STEM data. 
    The top two rows display the denoising results of ZSM-5 zeolite, while the last two rows showcase the denoising results of MOR zeolite. The orange and blue arrows point to Silicon and Oxygen atoms, respectively.
    * indicates that the dataset-based self-supervised denoising method was performed in a zero-shot learning manner. }
    \label{res_1z_mor}
\end{figure*}

\subsubsection{ZSM-5 \& MOR Zeolites STEM Data}
To further apply this method, two kinds of zeolites, ZSM-5 zeolite and MOR zeolite were imaged under very low-dose conditions and then denoised. 
It is worth mentioning that zeolites, as one kind of import nano-porous materials that were widely used in catalysis, separation, \textit{etc}, are extremely sensitive to electron beams. 
Thus, low electron dose conditions are applied, resulting in low SNR of STEM images which hinder the visualization of atomic structure. 

In Fig.~\ref{res_1z_mor}, we present the qualitative results obtained from denoising ZSM-5 and MOR zeolite STEM data using different methods. 
Conventional non-learning methods yield blurred denoising results that fail to capture the atomic structure clearly. 
Among the zero-shot denoising methods, S2S introduces artifacts resembling pepper salt noise, while NB2NB results in a blurred output due to artifacts.
In contrast, both FBI-D and Noise2SR clearly represent the atomic structure. However, in the zoomed-in region, FBI-D exhibits some structural artifacts, whereas Noise2SR demonstrates higher contrast with fewer artifacts.
In the Noise2SR result of ZSM-5 zeolite, not only the bright and shape contrast of Silicon (Si) atoms in 5-, 6- and 10- member ring (MR) can be well resolved, but also the weak contrast of light Oxygen (O) atom between two Si atoms can be clearly visualized. Similarly, in the Noise2SR result of MOR zeolite, the contrast of Si atoms can be clearly resolved while in the raw image, it is hard to identify because of the noise and low SNR. These results show the great potential of our Noise2SR method in denoising the HREM images, which helps to visualize and study the atomic structures in materials science.

\begin{figure}
    \centering
    \includegraphics[width=0.85\linewidth]{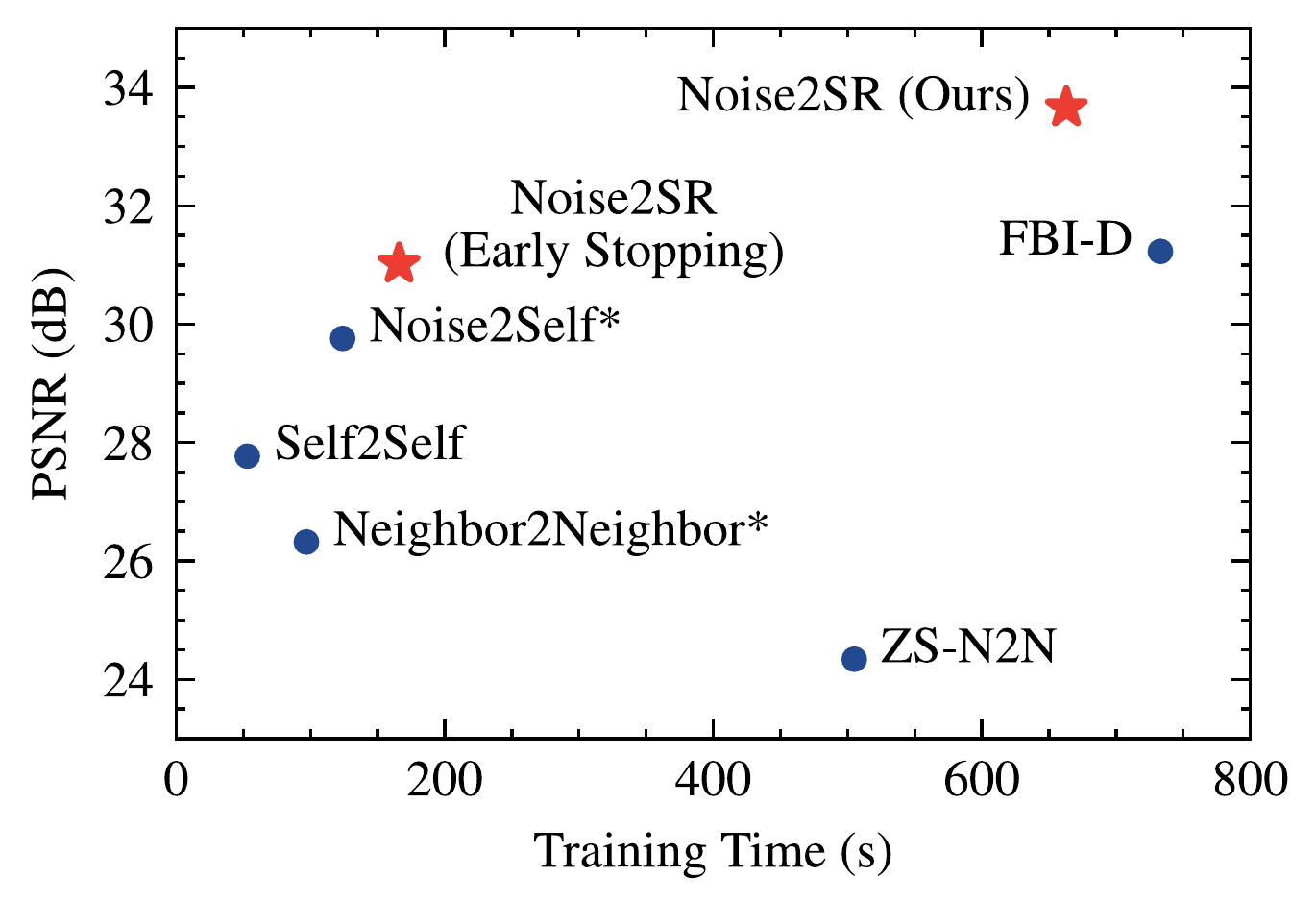}
    \caption{Training time of denoiser with zero-shot learning v.s the PSNR results on simulated TEM corrupted by Poisson-Gaussian noise with $(a = 0.05, b=0.02)$.}
    \label{fig:compared_time}
\end{figure}

\subsection{{Comparison of Time Required for Zero-shot Denoiser}}
 A zero-shot denoiser is an image-specific denoiser that requires training on the target noisy image. 
However, the excessive training time for zero-shot denoisers can limit their practical applications. 
In our experiments, we found that the training time of zero-shot denoiser takes several minutes, while the inference time is seconds, which can be neglected. 
We compared the training time required by different denoisers to achieve the performance results shown in Table~\ref{tab1}.
Additionally, we compared the performance of Noise2SR with early stopping,  which takes approximately the same training time as Noise2Self. 
To better illustrate the balance between training time and performance, we included Fig.~\ref{fig:compared_time}.
The results show that, compared to FBI-D, Noise2SR can achieve similar performance in less time. 
Furthermore, compared to Noise2Self, Noise2SR delivers better performance within the same training time. 
By extending the training time, Noise2SR can significantly outperform other methods.

\begin{figure}[!t]
    \centering
    \includegraphics[width = \columnwidth]{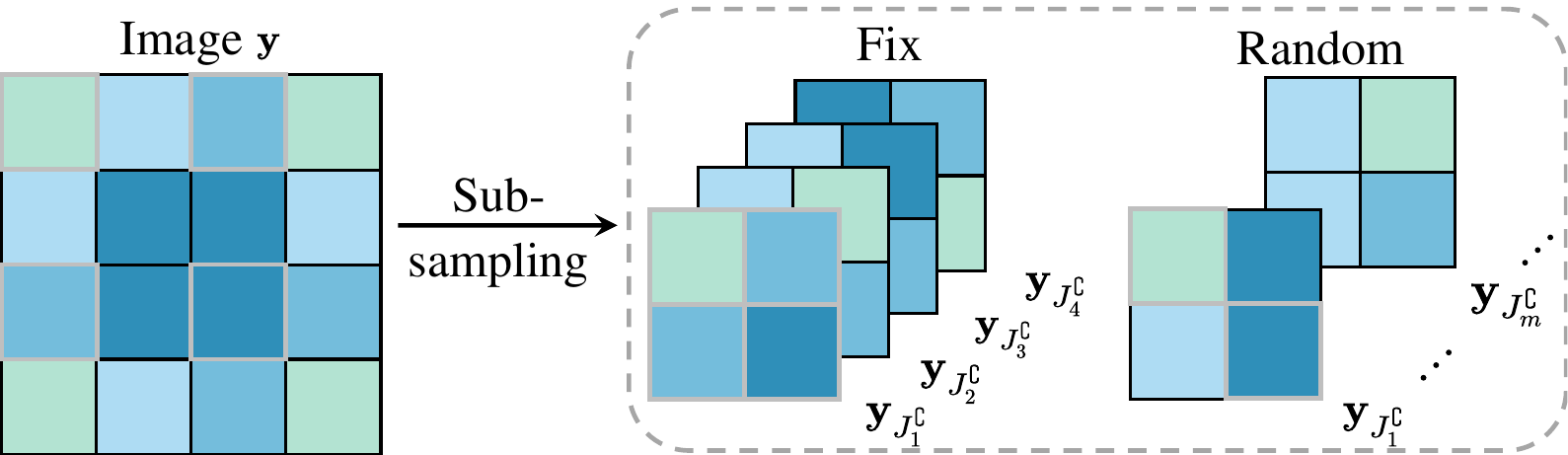}
    \caption{Comparison of fix-location and random location sub-sampling in Sub-sampler with sampling stride $(s = 2)$.}
    \label{fig:sub-sampling}
\end{figure}

\subsection{Effectiveness of Randomness in Sub-sampler}
\label{secIV-F}
\begin{figure}[!t]
    \centering
    \includegraphics[width=\columnwidth]{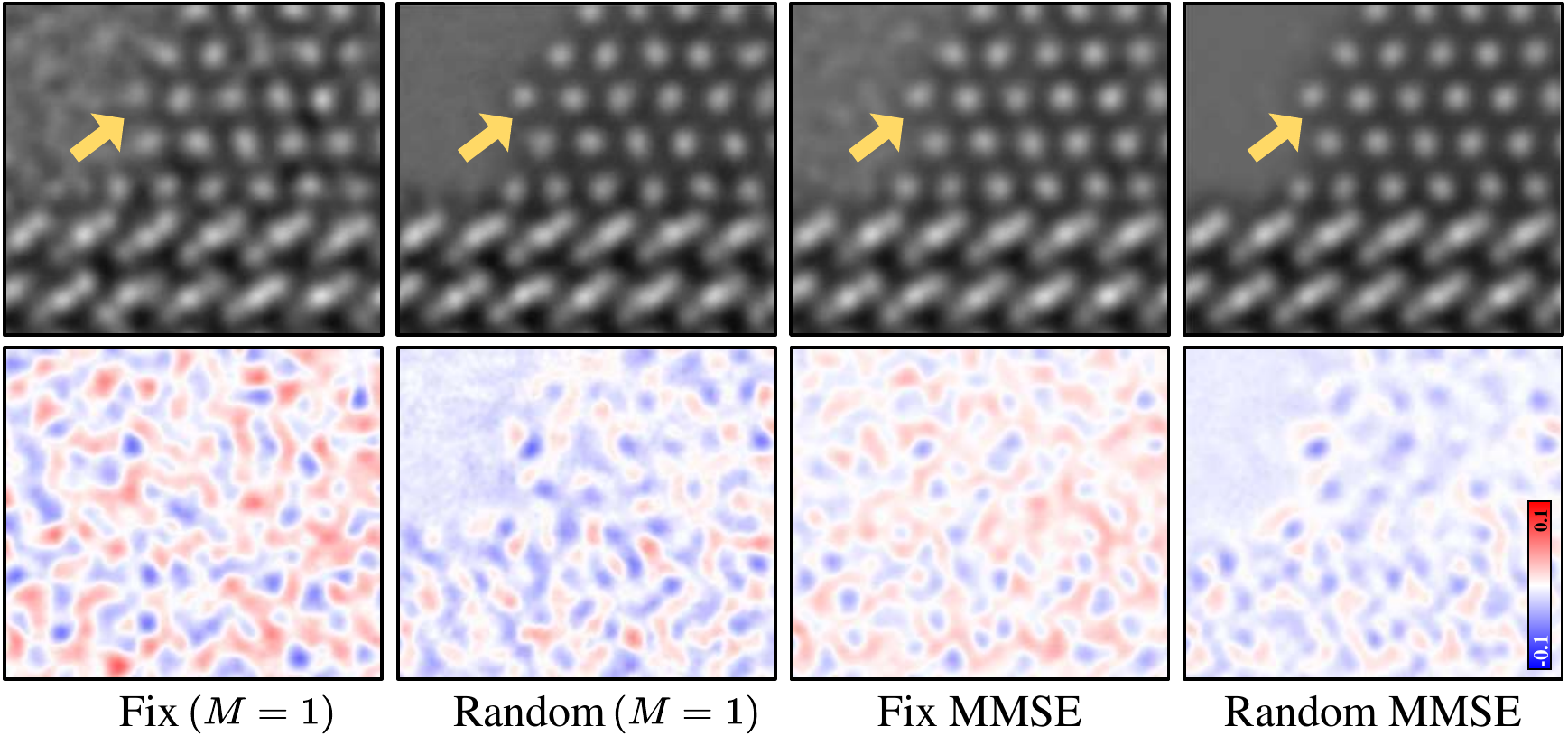}
    \caption{Qualitative results of fix and random location sub-sampling comparisons of denoising simulated TEM dataset.}
    \label{fig:res_sub_sampling}
\end{figure}

In this paper, we propose the Random Sub-sampler, a method for generating sub-sampled noisy images from full-size noisy images.
To evaluate the effectiveness of randomness in sub-sampler on the denoising performance of Noise2SR, we compare the fix-location and random location sampling strategies within sub-sampler.
Fig. \ref{fig:sub-sampling} illustrates the difference of fix and random location sub-sampling strategies of sampling stride $s = 2$. 
Table~\ref{tab:table2} and Fig.~\ref{fig:res_sub_sampling} show the quantitative and qualitative results, respectively. 
The results demonstrate that random sub-sampling improves the denoising performance compared to fixed location sub-sampling. 
Both the denoised results from an individual sub-sampled image $(M=1)$ and MMSE denoised results, random sub-sampling shows the superiority of fixed sub-sampling.
In the qualitative result, it is evident that the denoised images obtained from random sub-sampling exhibit clearer backgrounds and sharper atom contours compared to those obtained from fixed location sub-sampling. This observation is further supported by the error maps of the denoising results.

\begin{table}[!]
\centering
\caption{Comparisons of fix and random location sub-sampling strategy in sub-sampler. Quantitative results (PSNR/SSIM) are evaluated on simulated TEM datasets with three different noise levels. The best performance is highlighted in \textbf{bold}, while the second-best performance is \underline{underlined}.}
\label{tab:table2}
\resizebox{\columnwidth}{!}{
\begin{tabular}{lccc} 
\toprule
\textbf{Noise Parameters}               & $a=0.1$ & $a=0.05$ & $a=0.02$  \\ 
\midrule
\text{Fix} $(M=1)$    & 25.33/0.8732   & 27.45/0.9131    & 29.68/0.9401     \\
\text{Random} $(M=1)$  & 28.44/0.9235   & 30.16/0.9532    & 32.70/0.9729     \\
\text{Fix MMSE}     & \underline{29.36}/\underline{0.9443}   & \underline{31.92}/\underline{0.9655}    & \underline{34.57}/\underline{0.9788}     \\
\text{Random MMSE}  & \textbf{31.92/0.9645}   & \textbf{33.44/0.9793}    & \textbf{36.41/0.9880}    \\
\bottomrule
\end{tabular}
}
\end{table}

\subsection{{Ablation Studies}}

\label{secIV-E}
\subsubsection{Influence of Sampling stride in Random Sub-sampler}
To evaluate the influence of sampling stride $s$ on the denoising performance, we conducted experiments using the Noise2SR model with three different values of $s$. 
The experiments were performed on noisy simulated TEM datasets with corrupted Poisson-Gaussian noise ($a = 0.05, b = 0.02$).

\begin{figure}[!t]
    \centering
    \includegraphics[width = \columnwidth]{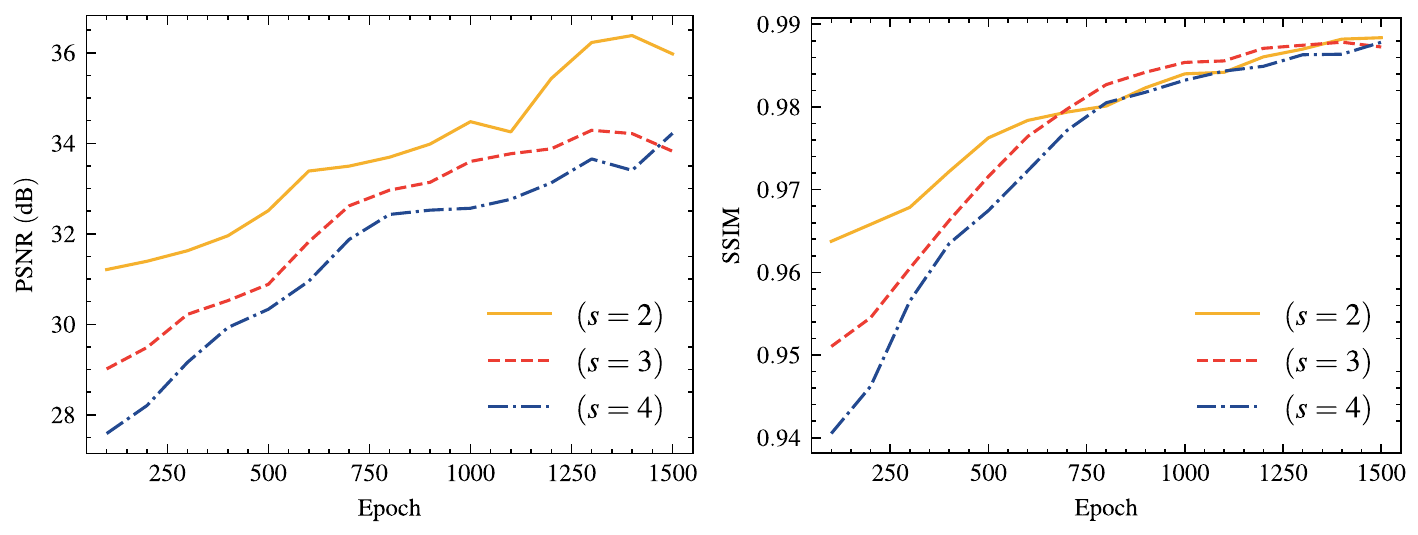}
    \caption{Performance curves of Noise2SR with different Random Sub-sampler parameters over training epochs on simulated TEM dataset with corrupted with Poisson-Gaussian noise ($a =0.05,b=0.02$).}
    \label{fig:res_fig5}
\end{figure}

\begin{figure}[!t]
    \centering
    \includegraphics[width = \columnwidth]{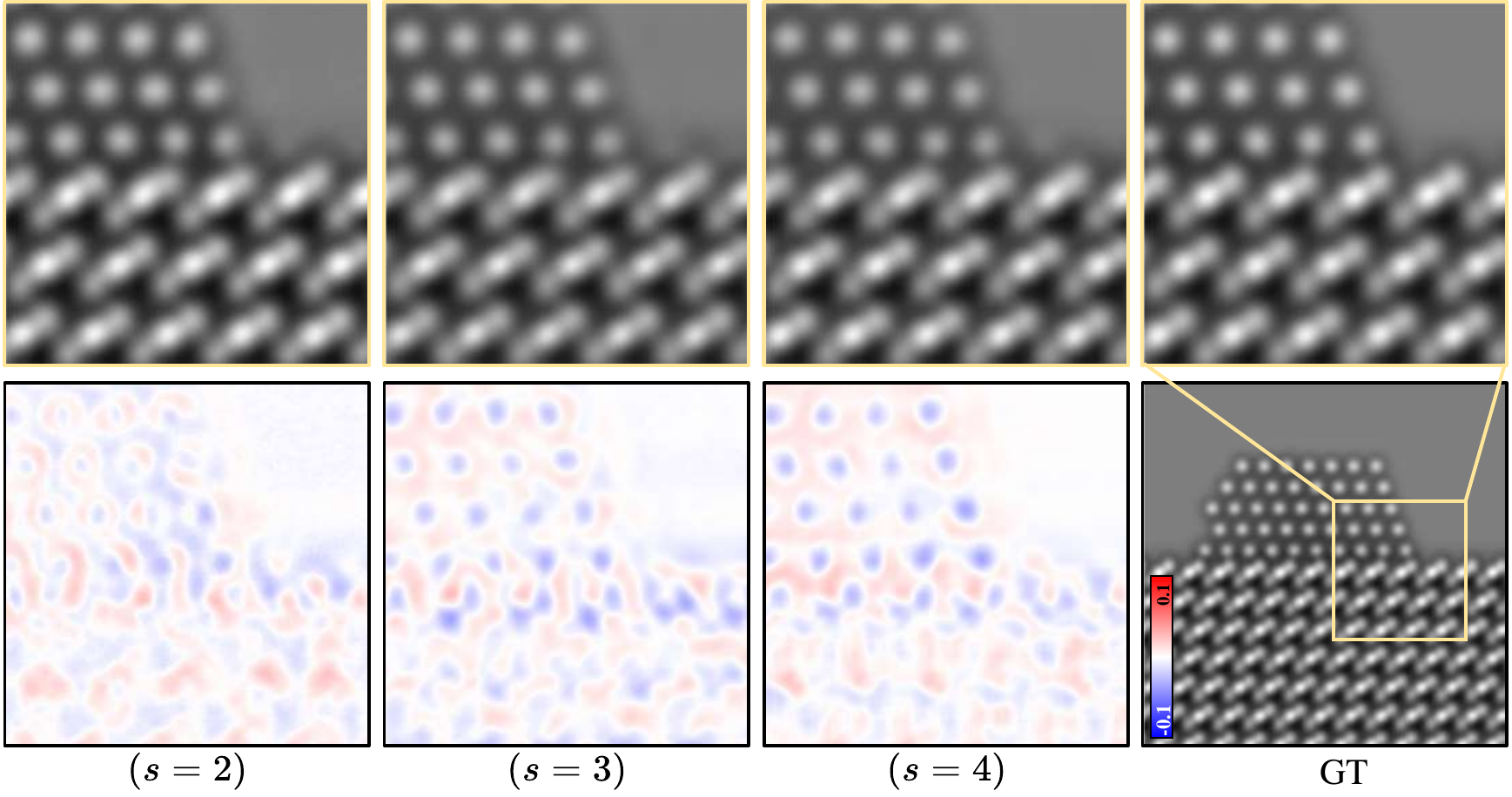}
    \caption{Qualitative results of three version of Noise2SR models with different Random Sub-sampler parameter of stride $(s = 2,3,4)$ on a test sample. }
    \label{fig:res_fig51}
\end{figure}

Fig.~\ref{fig:res_fig5}  presents the performing curves Noise2SR with different sampling stride $s$. 
The results show that Noise2SR with sampling stride $(s = 2)$ achieved the best denoising performance in PSNR/SSIM metrics.
Increasing the stride $s$ led to a decrease in denoising performance.
Fig.~\ref{fig:res_fig51}, shows qualitative comparisons of denoised results.
All the denoised results closely resembled the ground truth, but the error maps revealed a decrease in denoising performance as the stride $s$ increased. 

\subsubsection{Influence of the Sampling Times in Inference}
To assess the influence of the number of sub-sampling on the MMSE estimation of the clean signal during the inference stage, we compared the quality of the clean image estimation using different numbers of sub-sampling, ranging from 1 to 50. 
Fig.~\ref{res_fig6} showcases the variation in PSNR of the restored images as the number of sub-sampling increases. 
Furthermore, it presents a comparison of the quantitative and qualitative results of the MMSE estimation of clean images for different numbers of sub-sampling when $M =$ 1, 5, 20,  and 50.
From the figure, it is observed that as the number of samples increases, the PSNR of the estimated clean images also increases while the variance decreases. 
Notably, there is a significant improvement in the PSNR when the number of samples increases from 1 to 5, with an approximate gain of 2 dB (31.34 vs. 33.61). 
However, as the number of samples increases from 20 to 50, the improvement in the quality of the clean images becomes less significant. 
Additionally, the error maps of the qualitative results show that the difference between the estimated clean images for $M=20$ and $M=50$ is relatively small.
This observation is consistent with the qualitative results of the estimation of clean image for $M=20$ and $M=50$, where the difference between the two is relatively small.

\subsubsection{{Influence of Encoder Networks}}
{To further explore the influence of the encoder network on the performance of Noise2SR, we compare the U-net based encoder with two well-known transformer-based backbones Uformer~\cite{wang2022uformer} and Restormer~\cite{zamir2022restormer}.
To better demonstrate the superiority of Noise2SR, we also compare the performance of Noise2Self with the three encoders and a SOTA BSN, FBI-Net~\cite{byun2021fbi}.}
{Additionally, we compute the multiply-accumulate operations (MACs) of different networks to compare their computational complexity.}

\par In the experiments, we implement Uformer and Restormer following the same parameters as in the original papers. We evaluate their performance on two noisy samples with different PSNRs (3 dB and 10 dB) and compared their computational complexity. The quantitative results are shown in Table~\ref{tab:compared_encoder}.
In Noise2SR, the Uformer shows a slight performance degradation compared to the U-net-based encoder, while the Restormer shows a significant improvement of approximately 1.5 dB and 2 dB on two samples, respectively.
In contrast, the transformer-based encoders in Noise2Self do not exhibit a performance gain over the U-net encoder but rather a decrease in the SSIM evaluations.
For HREM images with high-structured features, CNN-based networks can achieve superior performance due to the inductive bias.
One possible explanation is that the masked-based strategy in Noise2Self is not suitable for $\mathcal{J}$-invariant denoisers training with transformer encoders. Since the self-attention is not $\mathcal{J}$-invariant, the transformer may overfit to mapping replacements to noisy pixels, resulting in performance degradation. Noise2SR enjoys more flexibility in terms of network structures for enhancing performance.
It is worth noting that the sub-sampling operation allows us to significantly reduce the computational complexity in the case of using a transformer-based encoder.

\begin{figure}
    \centering
    \includegraphics[width=\linewidth]{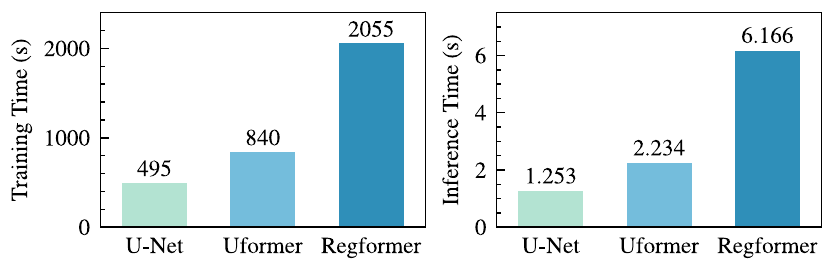}
    \caption{{Quantitative comparison of training and inference time of Noise2SR with three encoder networks. The inference time is computed on denoising an image with a size of $1024\times 1024$.}}
    \label{fig:encoder_time}
\end{figure}

\begin{figure}[!t]
    \centering
    \includegraphics[width=\columnwidth]{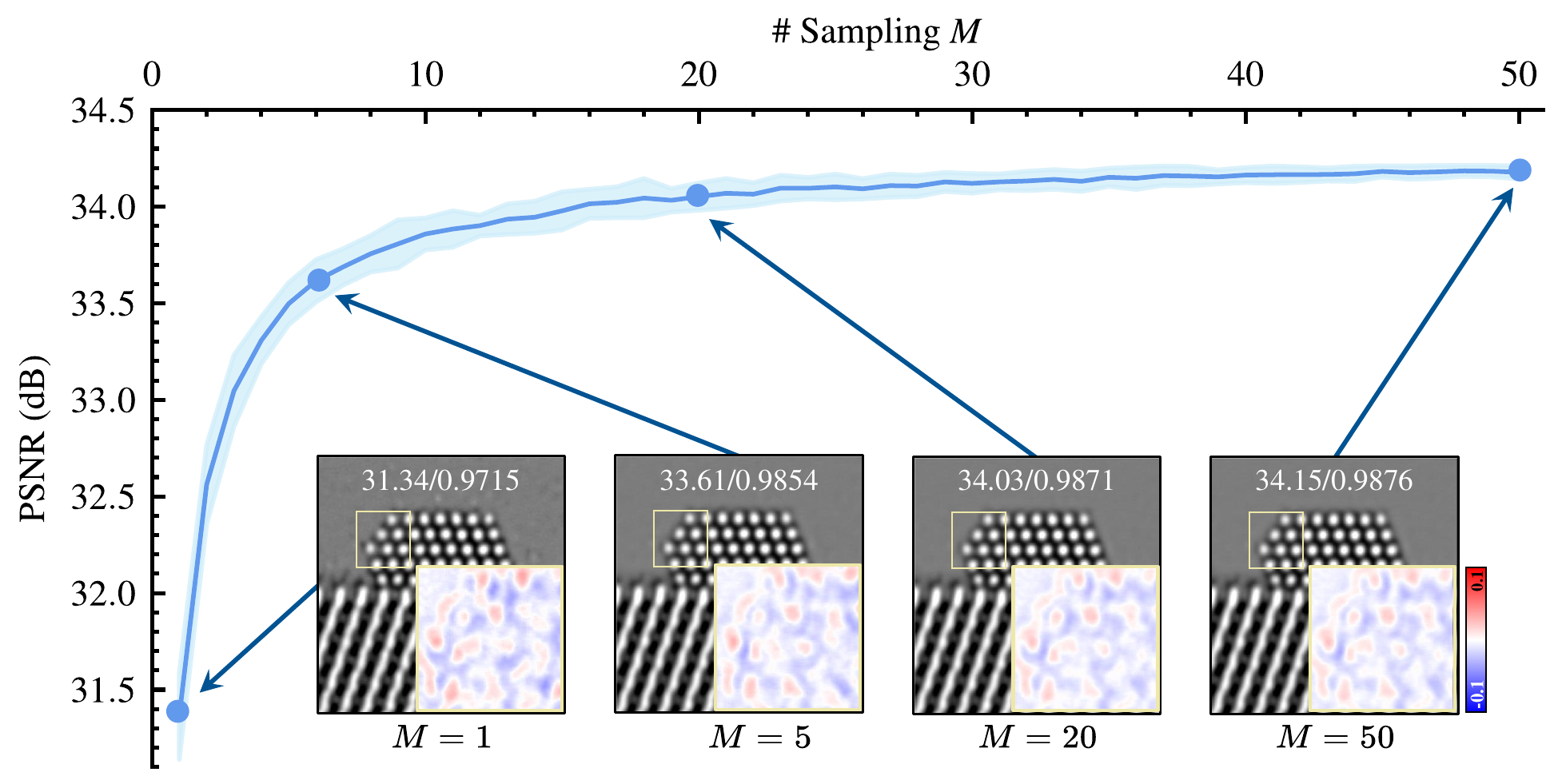}
    \caption{PSNR versus the number of sub-sampling times for MMSE estimation of the clean signal.}
    \label{res_fig6}
\end{figure}

{
Furthermore, we compared training and inference time for Noise2SR with three encoder networks. The quantitative comparison is shown in Fig.~\ref{fig:encoder_time}.
The results indicate that adopting a transformer-based encoder network increases training and inference time. Especially Regformer, it takes nearly five times longer than U-Net.
}

\begin{table}[!t]
\centering
\caption{Quantitative comparisons of Noise2Self and Noise2SR with three distinct encoder networks and FBI-Net on two noisy samples. The computational complexity of different networks is also compared. {Multiply-accumulate operations} (MACs) are computed on a patch size of $128\times 128 $. The best performance is highlighted in \textbf{bold}, while the second-best performance is \underline{underlined}. }
\label{tab:compared_encoder}
\resizebox{\columnwidth}{!}{
\begin{tabular}{clccc} 
\toprule
\textbf{SSL Strategy }               & \textbf{Encoder}   & \textbf{3 dB}          & \textbf{10 dB}         & \textbf{MACs (G)}  \\ 
\midrule
\multirow{3}{*}{Noise2Self} & \texttt{Unet}      & 25.20/0.7455 & 33.44/0.9306 &   \underline{4.61}    \\
                            & \texttt{Uformer}   & 25.56/0.6754 & 31.78/0.8899 &   10.24   \\
                            & \texttt{Restormer} & 26.17/0.6374 & 33.16/0.8665 &   35.21    \\ 
\midrule
\multirow{3}{*}{Noise2SR}   & \texttt{Unet}      & \underline{29.75}/\underline{0.9595} & \underline{35.89}/\underline{0.9881} &    6.82   \\
                            & \texttt{Uformer}   & 29.53/0.9563 & 34.98/0.9838 &    \textbf{4.15}   \\
                            & \texttt{Restormer} & \textbf{31.25/0.9630} & \textbf{37.84/0.9851} &    11.59   \\
\midrule
FBI-Net &  \texttt{FBI-Net} & 25.75/0.7434 & 34.33/0.9379 &  6.56\\
\bottomrule
\end{tabular}}
\end{table}

\subsubsection{Influence of Different Upsampling Strategies}
To evaluate the impact of different upsampling strategies on the denoising performance of Noise2SR,  three versions of the Noise2SR model with different upsampling techniques are compared: 1) Transposed convolution (TransConv.); 2) Pixel shuffling (PixelShuff.); 3) Bilinear interpolation (BiInter.). The performance of these models was evaluated by denoising two noisy samples with PSNR of 3 dB and 10 dB, respectively.

Table \ref{tab:table3} shows the quantitative results. 
The results show that for the test sample with a PSNR of 3 dB, the denoising performance of the three upsampling methods is comparable, with Noise2SR using pixel shuffling showing a slight advantage. 
For the test sample with a higher SNR of 10 dB, Noise2SR using transposed convolution achieves a slight improvement of 0.3 dB compared to the other two methods. 
{The results indicate that Noise2SR is relatively robust to various upsampling strategies, including bilinear interpolation, which has no training parameters. 
This flexibility allows for selecting upsampling strategies based on the desired trade-off between performance and speed in different application scenarios.}

\begin{table}[t]
\centering
\arrayrulecolor{black}
\centering
\caption{Ablation on different upsampling strategies in SR decoder. Quantitative results (PSNR/SSIM) are evaluated on two noisy samples with different SNRs in simulated TEM Dataset. The best performance is highlighted in \textbf{Bold}.}
\label{tab:table3}
\begin{tabular}{lccc} 
\toprule
 \textbf{Upsampling}              & \textbf{3 dB}                  & \textbf{10 dB }                    \\ 
\midrule
\texttt{TransConv.}      & 27.36/0.9292         & \textbf{36.22/0.9894}   \\
\texttt{PixelShuff.} & \textbf{27.51/0.9349} & 35.89/0.9881  \\
\texttt{BiInter.} & 27.31/0.9325  & 35.59/0.9876            \\
\bottomrule
\end{tabular}
\arrayrulecolor{black}
\end{table}

\subsection{Failure Cases Illustrations}
\begin{figure}
    \centering
    \includegraphics[width=\linewidth]{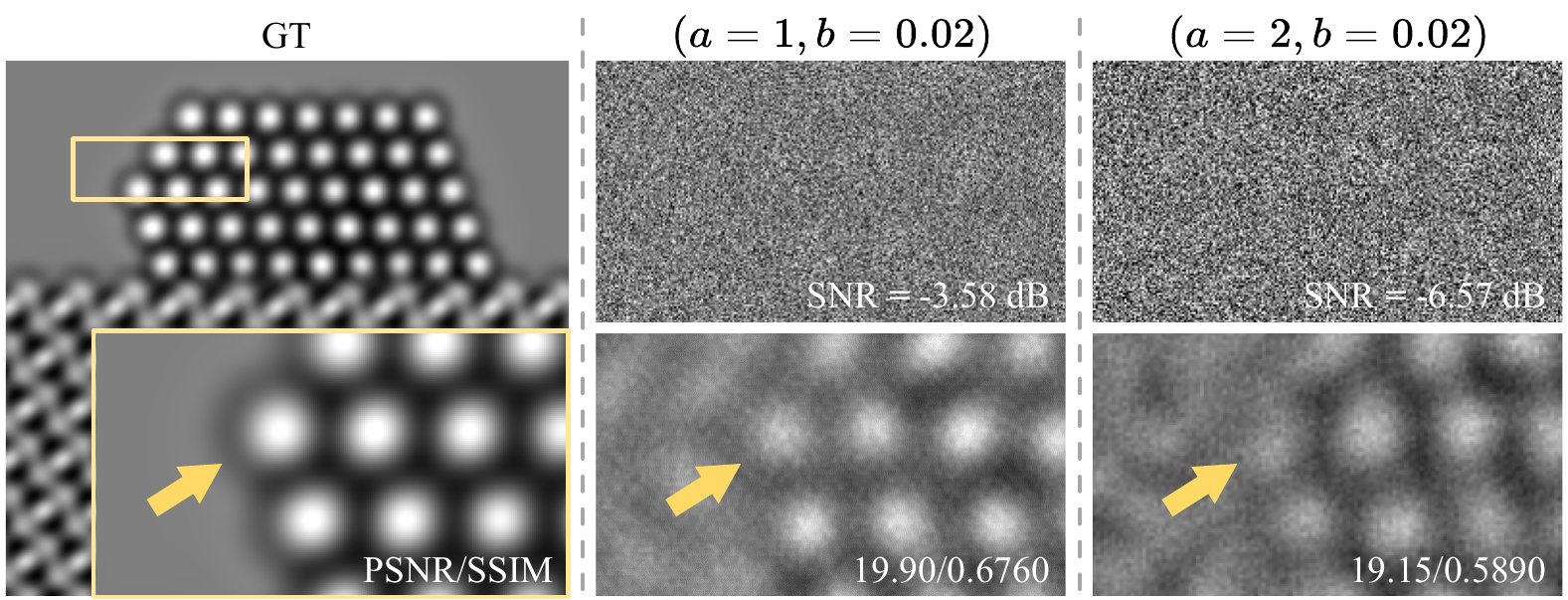}
    \caption{{Illustration of failure cases: quantitative and qualitative results of Noise2SR performance on denoising simulated noisy TEM image corrupted by Poisson-Gaussian noise with $(a=1, b= 0.02)$ and $(a=2, b= 0.02)$. }}
    \label{fig:failure_case}
\end{figure}

We present two failure cases of Noise2SR for a comprehensive evaluation and a better understanding of its limitations.
We test the performance of Noise2SR on denoising simulated TEM images that are extremely corrupted by severe Poisson-Gaussian noise with $(a=1, b= 0.02)$ and $(a=2, b= 0.02)$.
To more directly assess the severity of image corruption, we calculated the SNR of the noisy image using the following formula:
\begin{equation}
    \text{SNR}_\text{dB} = 10 \log_{10} \left ( \frac{\mathbb{E}[\mathbf{s}^2]}{\mathbb{E}[\mathbf{n}^2]} \right ),
\end{equation}
where $\mathbf{s}$ and $\mathbf{n}$ denote the clean signal and noise respectively.
Fig.~\ref{fig:failure_case} shows the failure cases.
Although some edges of the atoms are still recoverable in the denoised images, where almost no information is visible in the original noisy images, the background of the denoised results exhibits significant artifacts. Additionally, the outlines of the atoms, particularly at the edges, become very unclear, as highlighted by the yellow arrows.
{
We analyzed the potential reason for the failures could be: 
1) The severe Poisson-Gaussian noise violates the self-supervised denoising assumption that noise is uncorrelated with the signal;
2) the small local patches in the noisy image may lack sufficient statistical information for effective denoising. 
To address these issues, future work could explore novel sampling or noise transformation strategies to reduce the correlation between signal and noise. Additionally, designing a network with a larger receptive field or incorporating multi-scale sub-sampled noisy images might improve performance. }

\section{Conclusion \& Discussion}
In this work, we propose Noise2SR, zero-shot self-supervised learning (ZS-SSL) method for denoising HREM images.
We introduce the Random Sub-sampler to generate paired noisy images with different resolutions, preserving structural information while altering high-frequency signal-dependent noise. 
This enables effective noise removal through supervised training using paired noisy images.
By employing a super-resolved strategy, the network generates denoised results of the original image size from sub-sampled noisy inputs.
The generation of paired noisy images serves as data augmentation to mitigate overfitting in single image denoising.
Furthermore, we present an approximate MMSE estimation of the clean signal by ensembling multiple denoised results from randomly sub-sampled noisy images of a single input. This further enhances the denoising performance.
The training strategy and MMSE estimation enable Noise2SR to achieve superior denoising performance using only a single ultra-low SNR HREM image.
Experimental results on synthetic and real HREM data demonstrate that Noise2SR outperforms state-of-the-art ZS-SSL methods and achieves comparable performance to supervised methods. 
Despite achieving state-of-the-art performance in zero-shot self-supervised denoising for HREM images, Noise2SR has certain limitations.
One such limitation is that random sub-sampling can lead to a loss of high-frequency information in noisy input images, causing Noise2SR to produce oversmooth results in images with rich detail. 
Additionally, Noise2SR may fail to produce satisfactory results under extreme signal-correlated noise. 
Another limitation is that the zero-shot DL denoiser demands minutes of training, preventing it from achieving real-time denoising performance.
Despite these challenges, we believe that this work holds the potential to improve the signal-to-noise ratio in materials imaging domains.

\bibliographystyle{IEEEtran}
\bibliography{refs.bib}

\begin{thebibliography}{10}
\providecommand{\url}[1]{#1}
\csname url@samestyle\endcsname
\providecommand{\newblock}{\relax}
\providecommand{\bibinfo}[2]{#2}
\providecommand{\BIBentrySTDinterwordspacing}{\spaceskip=0pt\relax}
\providecommand{\BIBentryALTinterwordstretchfactor}{4}
\providecommand{\BIBentryALTinterwordspacing}{\spaceskip=\fontdimen2\font plus
\BIBentryALTinterwordstretchfactor\fontdimen3\font minus \fontdimen4\font\relax}
\providecommand{\BIBforeignlanguage}[2]{{%
\expandafter\ifx\csname l@#1\endcsname\relax
\typeout{** WARNING: IEEEtran.bst: No hyphenation pattern has been}%
\typeout{** loaded for the language `#1'. Using the pattern for}%
\typeout{** the default language instead.}%
\else
\language=\csname l@#1\endcsname
\fi
#2}}
\providecommand{\BIBdecl}{\relax}
\BIBdecl

\bibitem{o1978computed}
M.~O'keefe, P.~Buseck, and S.~Iijima, ``Computed crystal structure images for high resolution electron microscopy,'' \emph{Nature}, vol. 274, no. 5669, pp. 322--324, 1978.

\bibitem{spence1981experimental}
J.~C. Spence and A.~V. Crewe, ``Experimental high-resolution electron microscopy,'' 1981.

\bibitem{kisielowski2008detection}
C.~Kisielowski, B.~Freitag, M.~Bischoff, H.~Van~Lin, S.~Lazar, G.~Knippels, P.~Tiemeijer, M.~van~der Stam, S.~von Harrach, M.~Stekelenburg \emph{et~al.}, ``Detection of single atoms and buried defects in three dimensions by aberration-corrected electron microscope with 0.5-{\aa} information limit,'' \emph{Microscopy and Microanalysis}, vol.~14, no.~5, pp. 469--477, 2008.

\bibitem{kato2004reducing}
N.~I. Kato, ``Reducing focused ion beam damage to transmission electron microscopy samples,'' \emph{Journal of electron microscopy}, vol.~53, no.~5, pp. 451--458, 2004.

\bibitem{jiang2015electron}
N.~Jiang, ``Electron beam damage in oxides: a review,'' \emph{Reports on Progress in Physics}, vol.~79, no.~1, p. 016501, 2015.

\bibitem{faruqi2018direct}
A.~Faruqi and G.~McMullan, ``Direct imaging detectors for electron microscopy,'' \emph{Nuclear Instruments and Methods in Physics Research Section A: Accelerators, Spectrometers, Detectors and Associated Equipment}, vol. 878, pp. 180--190, 2018.

\bibitem{ercius20204d}
P.~Ercius, I.~Johnson, H.~Brown, P.~Pelz, S.-L. Hsu, B.~Draney, E.~Fong, A.~Goldschmidt, J.~Joseph, J.~Lee \emph{et~al.}, ``The 4d camera--an 87 khz frame-rate detector for counted 4d-stem experiments,'' \emph{Microscopy and Microanalysis}, vol.~26, no.~S2, pp. 1896--1897, 2020.

\bibitem{zhang2017beyond}
K.~Zhang, W.~Zuo, Y.~Chen, D.~Meng, and L.~Zhang, ``Beyond a gaussian denoiser: Residual learning of deep cnn for image denoising,'' \emph{IEEE Transactions on Image Processing}, vol.~26, no.~7, pp. 3142--3155, 2017.

\bibitem{tai2017memnet}
Y.~Tai, J.~Yang, X.~Liu, and C.~Xu, ``Memnet: A persistent memory network for image restoration,'' in \emph{Proceedings of the IEEE international conference on computer vision}, 2017, pp. 4539--4547.

\bibitem{zhang2018ffdnet}
K.~Zhang, W.~Zuo, and L.~Zhang, ``Ffdnet: Toward a fast and flexible solution for cnn-based image denoising,'' \emph{IEEE Transactions on Image Processing}, vol.~27, no.~9, pp. 4608--4622, 2018.

\bibitem{batson2019noise2self}
J.~Batson and L.~Royer, ``Noise2self: Blind denoising by self-supervision,'' in \emph{International Conference on Machine Learning}.\hskip 1em plus 0.5em minus 0.4em\relax PMLR, 2019, pp. 524--533.

\bibitem{krull2019noise2void}
A.~Krull, T.-O. Buchholz, and F.~Jug, ``Noise2void-learning denoising from single noisy images,'' in \emph{Proceedings of the IEEE/CVF conference on computer vision and pattern recognition}, 2019, pp. 2129--2137.

\bibitem{laine2019high}
S.~Laine, T.~Karras, J.~Lehtinen, and T.~Aila, ``High-quality self-supervised deep image denoising,'' \emph{Advances in Neural Information Processing Systems}, vol.~32, 2019.

\bibitem{cha2019fully}
S.~Cha and T.~Moon, ``Fully convolutional pixel adaptive image denoiser,'' in \emph{Proceedings of the IEEE/CVF International Conference on Computer Vision}, 2019, pp. 4160--4169.

\bibitem{quan2020self2self}
Y.~Quan, M.~Chen, T.~Pang, and H.~Ji, ``Self2self with dropout: Learning self-supervised denoising from single image,'' in \emph{Proceedings of the IEEE/CVF conference on computer vision and pattern recognition}, 2020, pp. 1890--1898.

\bibitem{xie2020noise2same}
Y.~Xie, Z.~Wang, and S.~Ji, ``Noise2same: Optimizing a self-supervised bound for image denoising,'' \emph{Advances in neural information processing systems}, vol.~33, pp. 20\,320--20\,330, 2020.

\bibitem{byun2021fbi}
J.~Byun, S.~Cha, and T.~Moon, ``Fbi-denoiser: Fast blind image denoiser for poisson-gaussian noise,'' in \emph{Proceedings of the IEEE/CVF Conference on Computer Vision and Pattern Recognition}, 2021, pp. 5768--5777.

\bibitem{moran2020noisier2noise}
N.~Moran, D.~Schmidt, Y.~Zhong, and P.~Coady, ``Noisier2noise: Learning to denoise from unpaired noisy data,'' in \emph{Proceedings of the IEEE/CVF Conference on Computer Vision and Pattern Recognition}, 2020, pp. 12\,064--12\,072.

\bibitem{huang2022neighbor2neighbor}
T.~Huang, S.~Li, X.~Jia, H.~Lu, and J.~Liu, ``Neighbor2neighbor: A self-supervised framework for deep image denoising,'' \emph{IEEE Transactions on Image Processing}, vol.~31, pp. 4023--4038, 2022.

\bibitem{pan2023random}
Y.~Pan, X.~Liu, X.~Liao, Y.~Cao, and C.~Ren, ``Random sub-samples generation for self-supervised real image denoising,'' in \emph{Proceedings of the IEEE/CVF International Conference on Computer Vision}, 2023, pp. 12\,150--12\,159.

\bibitem{wu2020unpaired}
X.~Wu, M.~Liu, Y.~Cao, D.~Ren, and W.~Zuo, ``Unpaired learning of deep image denoising,'' in \emph{European conference on computer vision}.\hskip 1em plus 0.5em minus 0.4em\relax Springer, 2020, pp. 352--368.

\bibitem{lequyer2022fast}
J.~Lequyer, R.~Philip, A.~Sharma, W.-H. Hsu, and L.~Pelletier, ``A fast blind zero-shot denoiser,'' \emph{Nature Machine Intelligence}, vol.~4, no.~11, pp. 953--963, 2022.

\bibitem{mansour2023zero}
Y.~Mansour and R.~Heckel, ``Zero-shot noise2noise: Efficient image denoising without any data,'' in \emph{Proceedings of the IEEE/CVF Conference on Computer Vision and Pattern Recognition}, 2023, pp. 14\,018--14\,027.

\bibitem{pang2021recorrupted}
T.~Pang, H.~Zheng, Y.~Quan, and H.~Ji, ``Recorrupted-to-recorrupted: Unsupervised deep learning for image denoising,'' in \emph{Proceedings of the IEEE/CVF conference on computer vision and pattern recognition}, 2021, pp. 2043--2052.

\bibitem{tian2022noise2sr}
X.~Tian, Q.~Wu, H.~Wei, and Y.~Zhang, ``Noise2sr: Learning to denoise from super-resolved single noisy fluorescence image,'' in \emph{Medical Image Computing and Computer Assisted Intervention--MICCAI 2022: 25th International Conference, Singapore, September 18--22, 2022, Proceedings, Part VI}.\hskip 1em plus 0.5em minus 0.4em\relax Springer, 2022, pp. 334--343.

\bibitem{lehtinen2018noise2noise}
J.~Lehtinen, J.~Munkberg, J.~Hasselgren, S.~Laine, T.~Karras, M.~Aittala, and T.~Aila, ``Noise2noise: Learning image restoration without clean data,'' in \emph{International Conference on Machine Learning}.\hskip 1em plus 0.5em minus 0.4em\relax PMLR, 2018, pp. 2965--2974.

\bibitem{lee2022ap}
W.~Lee, S.~Son, and K.~M. Lee, ``Ap-bsn: Self-supervised denoising for real-world images via asymmetric pd and blind-spot network,'' in \emph{Proceedings of the IEEE/CVF Conference on Computer Vision and Pattern Recognition}, 2022, pp. 17\,725--17\,734.

\bibitem{xu2020noisy}
J.~Xu, Y.~Huang, M.~Cheng, L.~Liu, F.~Zhu, Z.~Xu, and L.~Shao, ``Noisy-as-clean: Learning self-supervised denoising from corrupted image.'' \emph{IEEE Transactions on Image Processing: a Publication of the IEEE Signal Processing Society}, 2020.

\bibitem{tomasi1998bilateral}
C.~Tomasi and R.~Manduchi, ``Bilateral filtering for gray and color images,'' in \emph{Sixth international conference on computer vision (IEEE Cat. No. 98CH36271)}.\hskip 1em plus 0.5em minus 0.4em\relax IEEE, 1998, pp. 839--846.

\bibitem{pantelic2006discriminative}
R.~S. Pantelic, R.~Rothnagel, C.-Y. Huang, D.~Muller, D.~Woolford, M.~J. Landsberg, A.~McDowall, B.~Pailthorpe, P.~R. Young, J.~Banks \emph{et~al.}, ``The discriminative bilateral filter: an enhanced denoising filter for electron microscopy data,'' \emph{Journal of structural biology}, vol. 155, no.~3, pp. 395--408, 2006.

\bibitem{buades2005non}
A.~Buades, B.~Coll, and J.-M. Morel, ``A non-local algorithm for image denoising,'' in \emph{2005 IEEE computer society conference on computer vision and pattern recognition (CVPR'05)}, vol.~2.\hskip 1em plus 0.5em minus 0.4em\relax Ieee, 2005, pp. 60--65.

\bibitem{wei2010optimized}
D.-Y. Wei and C.-C. Yin, ``An optimized locally adaptive non-local means denoising filter for cryo-electron microscopy data,'' \emph{Journal of structural biology}, vol. 172, no.~3, pp. 211--218, 2010.

\bibitem{dabov2007bm3d}
K.~Dabov, A.~Foi, V.~Katkovnik, and K.~Egiazarian, ``Image denoising by sparse 3-d transform-domain collaborative filtering,'' \emph{IEEE Transactions on Image Processing}, vol.~16, no.~8, pp. 2080--2095, 2007.

\bibitem{ede2021deep}
J.~M. Ede, ``Deep learning in electron microscopy,'' \emph{Machine Learning: Science and Technology}, vol.~2, no.~1, p. 011004, 2021.

\bibitem{kalinin2022machine}
S.~V. Kalinin, C.~Ophus, P.~M. Voyles, R.~Erni, D.~Kepaptsoglou, V.~Grillo, A.~R. Lupini, M.~P. Oxley, E.~Schwenker, M.~K. Chan \emph{et~al.}, ``Machine learning in scanning transmission electron microscopy,'' \emph{Nature Reviews Methods Primers}, vol.~2, no.~1, p.~11, 2022.

\bibitem{quan2019removing}
T.~M. Quan, D.~G.~C. Hildebrand, K.~Lee, L.~A. Thomas, A.~T. Kuan, W.-C.~A. Lee, and W.-K. Jeong, ``Removing imaging artifacts in electron microscopy using an asymmetrically cyclic adversarial network without paired training data,'' in \emph{2019 IEEE/CVF International Conference on Computer Vision Workshop (ICCVW)}.\hskip 1em plus 0.5em minus 0.4em\relax IEEE, 2019, pp. 3804--3813.

\bibitem{wang2020noise2atom}
F.~Wang, T.~R. Henninen, D.~Keller, and R.~Erni, ``Noise2atom: unsupervised denoising for scanning transmission electron microscopy images,'' \emph{Applied Microscopy}, vol.~50, no.~1, pp. 1--9, 2020.

\bibitem{chong2023m}
X.~Chong, M.~Cheng, W.~Fan, Q.~Li, and H.~Leung, ``M-denoiser: Unsupervised image denoising for real-world optical and electron microscopy data,'' \emph{Computers in Biology and Medicine}, vol. 164, p. 107308, 2023.

\bibitem{lin2021temimagenet}
R.~Lin, R.~Zhang, C.~Wang, X.-Q. Yang, and H.~L. Xin, ``Temimagenet training library and atomsegnet deep-learning models for high-precision atom segmentation, localization, denoising, and deblurring of atomic-resolution images,'' \emph{Scientific reports}, vol.~11, no.~1, p. 5386, 2021.

\bibitem{mohan2022deep}
S.~Mohan, R.~Manzorro, J.~L. Vincent, B.~Tang, D.~Y. Sheth, E.~P. Simoncelli, D.~S. Matteson, P.~A. Crozier, and C.~Fernandez-Granda, ``Deep denoising for scientific discovery: A case study in electron microscopy,'' \emph{IEEE Transactions on Computational Imaging}, vol.~8, pp. 585--597, 2022.

\bibitem{zhou2020awgn}
Y.~Zhou, J.~Jiao, H.~Huang, Y.~Wang, J.~Wang, H.~Shi, and T.~Huang, ``When awgn-based denoiser meets real noises,'' in \emph{Proceedings of the AAAI Conference on Artificial Intelligence}, vol.~34, no.~07, 2020, pp. 13\,074--13\,081.

\bibitem{shi2016real}
W.~Shi, J.~Caballero, F.~Husz{\'a}r, J.~Totz, A.~P. Aitken, R.~Bishop, D.~Rueckert, and Z.~Wang, ``Real-time single image and video super-resolution using an efficient sub-pixel convolutional neural network,'' in \emph{Proceedings of the IEEE conference on computer vision and pattern recognition}, 2016, pp. 1874--1883.

\bibitem{lim1990awf}
J.~S. Lim, ``Two-dimensional signal and image processing,'' \emph{Englewood Cliffs}, 1990.

\bibitem{kingma2014adam}
\BIBentryALTinterwordspacing
D.~P. Kingma and J.~Ba, ``Adam: A method for stochastic optimization,'' in \emph{ICLR (Poster)}, 2015. [Online]. Available: \url{http://arxiv.org/abs/1412.6980}
\BIBentrySTDinterwordspacing

\bibitem{wang2004image}
Z.~Wang, A.~C. Bovik, H.~R. Sheikh, and E.~P. Simoncelli, ``Image quality assessment: from error visibility to structural similarity,'' \emph{IEEE transactions on Image Processing}, vol.~13, no.~4, pp. 600--612, 2004.

\bibitem{boyat2015review}
A.~K. Boyat and B.~K. Joshi, ``A review paper: Noise models in digital image processing,'' \emph{Signal \& Image Processing}, vol.~6, no.~2, p.~63, 2015.

\bibitem{sarder2006deconvolution}
P.~Sarder and A.~Nehorai, ``Deconvolution methods for 3-d fluorescence microscopy images,'' \emph{IEEE signal processing magazine}, vol.~23, no.~3, pp. 32--45, 2006.

\bibitem{meiniel2018denoising}
W.~Meiniel, J.-C. Olivo-Marin, and E.~D. Angelini, ``Denoising of microscopy images: a review of the state-of-the-art, and a new sparsity-based method,'' \emph{IEEE Transactions on Image Processing}, vol.~27, no.~8, pp. 3842--3856, 2018.

\bibitem{bahler2022pogain}
N.~B{\"a}hler, M.~El~Helou, {\'E}.~Objois, K.~Okumu{\c{s}}, and S.~S{\"u}sstrunk, ``Pogain: Poisson-gaussian image noise modeling from paired samples,'' \emph{IEEE Signal Processing Letters}, vol.~29, pp. 2602--2606, 2022.

\bibitem{wang2022uformer}
Z.~Wang, X.~Cun, J.~Bao, W.~Zhou, J.~Liu, and H.~Li, ``Uformer: A general u-shaped transformer for image restoration,'' in \emph{Proceedings of the IEEE/CVF conference on computer vision and pattern recognition}, 2022, pp. 17\,683--17\,693.

\bibitem{zamir2022restormer}
S.~W. Zamir, A.~Arora, S.~Khan, M.~Hayat, F.~S. Khan, and M.-H. Yang, ``Restormer: Efficient transformer for high-resolution image restoration,'' in \emph{Proceedings of the IEEE/CVF conference on computer vision and pattern recognition}, 2022, pp. 5728--5739.

\end{thebibliography}

\end{document}